\documentclass[prb,amsmath,amssymb,twocolumn,aps,showpacs]{revtex4-1}
\usepackage{graphicx}
\usepackage{dcolumn}
\usepackage{bm}
\usepackage{braket}
\usepackage{color}
\begin{document}

\title{Tuning edge state localization in graphene nanoribbons by in-plane bending}
\author{S.G. Stuij$^1$}
\author{P. H. Jacobse$^2$}
\author{V. Juri\v{c}i\'c$^1$}
\author{C. Morais Smith$^1$}
\affiliation{
$^1$Institute for Theoretical Physics, Center for Extreme Matter and Emergent Phenomena, Utrecht University, Leuvenlaan $4$, $3584$ CE Utrecht, the Netherlands \\
$^2$Debye Institute for Nanomaterials Science, Condensed Matter and Interfaces, 
Utrecht University, P.O. Box $80.000$, $3508$ TA Utrecht, the Netherlands}

\date{\today}

\begin{abstract}
The electronic properties of graphene are influenced by both geometric confinement and strain. We study the electronic structure of in-plane bent graphene nanoribbons, systems where confinement and strain are combined. To understand its electronic properties, we develop a tight-binding model that has a small computational cost and is based on exponentially decaying hopping and overlap parameters. Using this model, we show that the edge states in zigzag graphene nanoribbons are sensitive to bending and develop an effective dispersion that can be described by a one-dimensional atomic chain model. Because the velocity of the electrons at the edge is proportional to the slope of the dispersion, the edge states become gradually delocalized upon increasing the strength of bending.
\end{abstract}


\maketitle

\section{\label{sec:level1}Introduction}
Many of graphene's remarkable features stem from two facts. The first is that its low energy quasiparticles are linearly dispersive and can be effectively described as Dirac fermions;\cite{CastroNeto2009} the second is that graphene is a two-dimensional ultrathin membrane that holds promises to revolutionize the current nanotechnology.\cite{Geim2007} In addition, this 2D membrane can be cut into 1D structures, so-called graphene nanoribbons (GNRs), which exhibit different transport properties, depending on their termination. Armchair terminated GNRs are usually gapped and therefore insulating. By virtue of their band gap, they can be used to create field-effect transistors.\cite{Wang2008,Bennett2013} On the other hand, zigzag GNRs (ZGNRs) show localized edge states that may be spin-polarized.\cite{Fujita1996} Although armchair-type GNRs have been successfully synthesised using bottom up approaches,\cite{Cai2010,Chen2013,Narita2014,Basagni2015} ZGNRs still remain elusive. Recently, patterned graphene with zigzag edges\cite{Shi2011} and GNRs with mixed armchair and zigzag terminations extending through a few lattice constants\cite{Han2014} have been experimentally realized. Despite the fact that the edges are not completely of zigzag type, they turned out to be of sufficient quality to confirm the prediction that the edge states become magnetized.\cite{Magda2014, Son2006}

Besides the geometrical confinement, another research area that has attracted much attention recently is the study of elastic deformations in graphene. Interest in this topic originated mainly from the theoretical prediction that strain may couple to the Dirac fermions as a pseudo-magnetic field (a magnetic field that preserves time-reversal symmetry). The subject was initially studied in the light of deformations in carbon nanotubes.\cite{Kane1997} After the rise of graphene, this research direction grew in prominence by the vision of using strain as a way to tune graphene's properties and use it in developing an all-graphene electronics.\cite{Amorim2015} The pursuit of strain engineering\cite{Pereira2009a} was pioneered by the experimental observation of ``pseudo"-Landau levels in strained graphene,\cite{Levy2010} and has been recently corroborated by fascinating examples of graphene spirals.\cite{Avdoshenko2013,Korhonen2014,Zhang2014}

In this paper, we study graphene systems that are both geometrically confined and strained, thus combining the two research areas through a specific example: in-plane bent GNRs. These systems have been theoretically investigated using  a model based on density-functional theory.\cite{Koskinen2012} In addition, they have been proposed as a  graphene geometry where strain couples as a uniform pseudo-magnetic field.\cite{Guinea2010} Recently, these systems have been experimentally realized by pushing a GNR with the tip of a scanning tunneling microscope.\cite{Lit2015} Although the experimentally synthesized bent samples are armchair terminated, here we concentrate on in-plane bent ZGNRs and study the dependence of the electronic behavior on the bending angle. We furthermore investigate the dependence of the electronic structure on the type of bending. Our studies complement recent investigations of the mechanical properties of these systems.\cite{Qi2014}

As a main result, we find that the bending leads to an increased dispersion in the otherwise almost flat edge states. The bending breaks the symmetry between the inner and the outer edges, causing an effective compression of the inside and elongation of the outside edge. To distinguish the contribution of each edge state to the dispersion, we compare our findings to straight ZGNRs under compressive or tensile strain. Our results show that by tuning the bending angle, the edge states become dispersive and hence delocalized.

From a more abstract perspective, we can view the deformations in graphene in terms of Japanese paper-art. Within this analogy, straight GNRs emerge as the art of \emph{paper cutting} graphene. On the other hand, \emph{origami}, the traditional Japanese art of \emph{paper folding}, is connected to the study of strain in graphene. These two come together in graphene \emph{kirigami},\cite{Qi2014} in which cutting and folding are combined. Here, the cutting refers to the specific termination of the GNR, as well as to the fact that the hexagonal unit cells are empty (cut), and can thus be deformed in a variety of ways. A bent GNR is a very specific and not so complicated type of graphene \emph{kirigami}, but precisely due to its relative simplicity, it is possible to study its electronic properties in depth. For this reason, the system is a good probe to understand how the electronic behavior arising from confinement and termination is affected by strain. Therefore, we may generalize the notion of \emph{kirigami} to more complicated graphene nanostructures, and apply a similar approach to understand their electronic properties. Knowing what is to be expected in this simple case, may help us understand more complicated situations.

This paper is organised as follows. In section \ref{sec:model} we introduce a tight-binding model with exponentially decaying hopping and overlap parameters, that we argue is suitable to study confined strained graphene systems. To the best of our knowledge, this particular tight-binding model has not been used previously to study GNRs, but turns out to capture all the relevant features of the band structure. We then introduce two types of bending, which allow us to optimize the computational cost. In section \ref{sec:results}, we apply our model to study the effects of bending on the edge state and find that their localization can be tuned by bending. Conclusions are provided in section \ref{sec:conclusions}.

\section{\label{sec:model}A minimal tight-binding model for bent GNRs}
\subsection{Three-parameter tight-binding model for strained confined graphene systems}
The electronic structure of graphene is usually derived using a tight-binding model with one $p_z$-orbital per site. 
If we assume a graphene system with $n$ sites positioned at $\mathbf{r}_i$, the single-electron wavefunction is given by
\begin{equation}\label{eq:TBwave}
\ket{\psi}=\sum_{i=1}^n c_i \ket{\phi_i}.
\end{equation}
Here, $\ket{\phi_i}$ are in general site-dependent basis states, which are assumed to be normalized. The vector $\mathbf{c}=(c_1,\ldots                           ,c_n)^T$ thus completely specifies the electron state. The Schr\"odinger equation can then be reduced to the $n\times n$ matrix equation
\begin{equation}\label{eq:TBmatrix}
(S\mathcal{E} +T)\mathbf{c}=E S \mathbf{c},
\end{equation}
where $E$ is the energy associated with the state specified by $\mathbf{c}$. Here, we have split the Hamiltonian matrix $H$, the elements of which are given by $H_{ij}=\braket{\phi_i|\hat{H}|\phi_j}$, into the so-called hopping matrix ($T$), the diagonal on-site energy matrix ($\mathcal{E}$), and the overlap matrix ($S$), such that  $H=S\mathcal{E}+T$. The elements of the overlap matrix are given by $S_{ij}=\braket{\phi_i|\phi_j}$. The matrix $\mathcal{E}$ is diagonal, with the elements corresponding to on-site energies, $\mathcal{E}\ket{\phi_j}=\epsilon_j\ket{\phi_j}$, as also defined in Ref.\ \onlinecite{Goerbig2011}.
Note that a more standard convention defines the on-site energy as the expectation value of the energy in a certain state and the hopping matrix as non-diagonal elements of the Hamiltonian matrix. However, the convention used here allows us to treat the on-site energy as a simple shift in $E$ if the on-site energy is the same for each state.

In general, we now have $n(n+1)$ parameters, the elements of the matrices. In tight-binding, these parameters may be found by fitting to a reference calculation, rather than calculating them explicitly as integrals over basis functions. However, a model with $n(n+1)$ parameters is impossible to fit when $n$ is not very small. Therefore, additional assumptions are made in order to reduce the parameter space. In graphene, translational symmetry allows one to use periodic boundary conditions. Since there is no longer a difference between individual sites, the on-site energy, hopping, and overlap parameters become site independent.

A common procedure is to consider a two-parameter model that only takes nearest-neighbor (NN) and next-nearest-neighbor (NNN) hopping into account, and  assumes orthogonal basis states. In this case, the site-independent on-site energy $\epsilon_0$ is left unspecified, as it leaves the eigenvectors invariant and produces only an absolute shift in the  spectrum.\cite{CastroNeto2009} However, we are interested in a model for the graphene system that can describe a bent GNR.  For such a model, we have to specify the dependence of the hopping and overlap parameters on the distance, and, at the same time, the parameters of the model should not change when the system is geometrically confined, e.g. when graphene is confined to a GNR. This last condition would allow us to fit the parameters to a graphene reference calculation and not to a reference calculation for the specific GNR we study. We find that instead of the usual convention, a non-orthogonal model better satisfies these two conditions. First, we introduce the model and later argue why it compares positively to an orthogonal model.

The tight-binding model we use is based on non-orthogonal site-independent basis states, which in real space are given by $\braket{\mathbf{r}|\phi_i}=\phi(\mathbf{r}-\mathbf{r}_i)$. Next to that, we assume that the hopping and overlap parameters between these states are such that $t_{ij}=t(\mathbf{r}_j-\mathbf{r}_i)$ and $S_{ij}=s(\mathbf{r}_j-\mathbf{r}_i)$ where $s,t$ are exponentially decaying functions, given by
\begin{equation}\label{eq:exppar}
\begin{split}
    t(\mathbf{r}) = \begin{cases}
               t_0 e^{\kappa(1 - |r|/a)}               & |r|>0\\
               0                                         & |r|=0\\
           \end{cases},
           \\
   s(\mathbf{r}) = \begin{cases}
   s_0 e^{\kappa(1 - |r|/a)}               & |r|>0\\
   1                                         & |r|=0\\
\end{cases}.
\end{split}
\end{equation}
Here, $a$ is the NN distance of graphene and $t_0$ and $s_0$ are the values of the NN-hopping and overlap parameter, respectively. Note that the on-site hopping parameter is zero and that the overlap of an orbital with itself is one. The dimensionless constant $\kappa$ determines the fall-off rate of the hopping. Although this procedure introduces a discontinuity in the overlap that cannot be physically realistic, we will assume that the strain sizes are small enough, such that this effect can be neglected. We further assume that the hopping and overlap parameters are proportional to each other, which implies that the parameter $\kappa$ is the same for both.

This model satisfies the first condition we mentioned, a dependence of the hopping and overlap parameters on the distance, better than an orthogonal model. This can be seen by noting that in studies of strained graphene, exponentially decaying functions have been used for parameters corresponding to orthogonal basis states. \cite{Pereira2009b,Ribeiro2009} However, efforts to reproduce the asymmetric band structure of graphene using up to $20$ fitted hoppings have resulted in subsequent parameters sometimes having opposite signs and clearly not following a trend that can be described with an exponential decay.\cite{Jung2013} On the other hand, if we relax the orthogonality condition, hopping and overlap are approximately exponentially decaying.\cite{Reich2002} When overlap is ignored in our parametrization ($s_0=0$), the model would be reduced to the one used in Ref. \onlinecite{Pereira2009b}. Such a model does not reproduce the correct particle-hole asymmetry. Nevertheless, for low energies the overlap becomes less important and it would yield a good estimate of the spectrum. Orthogonal models which involve a non-exponential dependence on distance have also been used. Ref. \onlinecite{Koskinen2012}, for instance, introduces a separate linear dependence for both the NN and NNN hopping. One reason why this model is disadvantageous is that it has four fitting parameters instead of three, as in our case.

An even more important reason for adopting the non-orthogonal approach is that these parameters are less dependent on the specific confinement than orthogonal parameters, thus better satisfying the second condition. To understand this, we note that in a quantum-confined graphene system we cannot expect all the hopping parameters to have the same value as the bulk parameters, since now the edge needs to be taken into account. For orthogonal states this is due, in part, to the fact that these states are a linear combination of $p_z$-orbitals obtained using an orthogonalization scheme, like the L\"owdin one.\cite{Aiken1980} These states are not the same on the edge and in the bulk, which also results  in a difference of on-site energy and hopping between bulk and edge. Therefore, it is more realistic to assume non-orthogonal basis states for the tight-binding model. This allows us to get the parameters from fitting to a graphene reference calculation and then apply it to the specific confined structure in which we are interested. A model based on nonorthogonal-basis states would be more universal than an orthogonal one for that reason. In Ref.~ \onlinecite{Koskinen2012}, an orthogonal tight-binding model is used and indeed we see that different hopping values are assumed for different GNRs: NNN hopping is zero for AGNRs and non-zero for ZGNRs. A more precise way to treat the edge effect requires the introduction of a different hopping at the edge.\cite{Sasaki2009,Hancock2010} However, for the sake of simplicity, we  neglect this effect here.

\begin{figure}[tb]
\centering
 \includegraphics[width=0.45\textwidth]{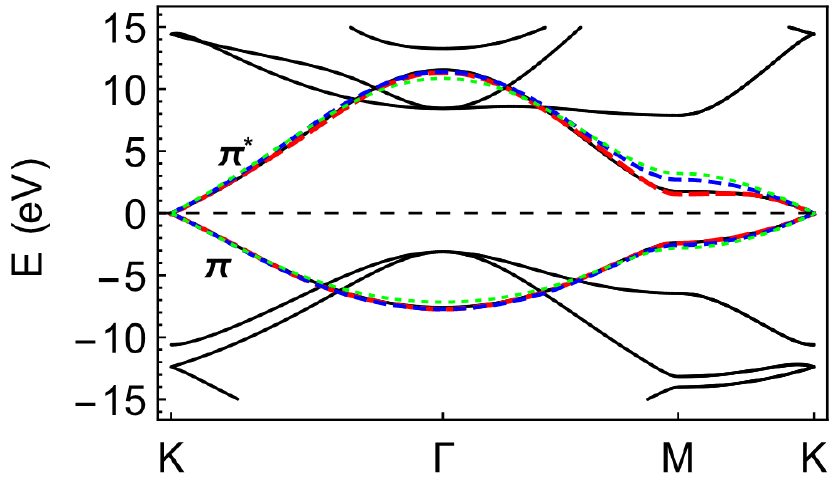}
    \caption{(Color online.) Plots of the dispersion relation for graphene along the line connecting the $\Gamma-M-K-\Gamma$ points of the Brillouin zone. The green-dashed curve corresponds to the two-parameter orthogonal dispersion of Ref.~\onlinecite{CastroNeto2009}, where the NN and NNN hopping parameters are $t=-3.00236$ eV and $t'=0.20509$ eV, and $\epsilon_0$ has been chosen such that the $K$ points are at zero energy. The blue-dashed curve depicts the orbital dispersion with exponentially decaying hopping and overlap parameter, described by Eq.~(\ref{eq:exppar}), with values $t_0=-2.8$ eV, $s_0=0.2$ , $\kappa=2.6$, and $\epsilon_0 = -1.28$ eV, chosen such that the zero energy is at the $K$ points. The red-dashed curve  corresponds to an orthogonal-basis dispersion taking into account the first $15$ hoppings,\cite{Jung2013} and the black-solid line corresponds to DFT calculations made using the QuantumWise software and a H\"uckel type basis set.\cite{Brandbyge2002}}\label{fig:graphenedispersion}
\end{figure}

We have argued that the parameters of the model can be obtained by fitting to a reference graphene spectrum. In the periodic graphene case, Bloch's theorem is used to reduce Eq.~(\ref{eq:TBmatrix}) to a $2\times2$ matrix equation, with wave functions labeled by the wavevector $\mathbf{k}$ in the Brillouin zone of graphene. In that case, the solution of this equation is equivalent to the one described in Ref.~\onlinecite{Reich2002}. By fitting to a reference first-principle spectrum, we find that $t_0=-2.8$ eV, $s_0=0.2$, and $\kappa=2.6$ gives a reasonable match, which is also not very far off from the parameters used in Ref.~\onlinecite{Reich2002}. Although a more elaborated fitting method would allow us to find parameters that reproduce the reference spectrum more closely, we settle with these because we are mostly interested in global features and not in extremely precise quantitative results.

The dispersion of graphene along a line connecting high-symmetry points of the Brillouin zone is shown in Fig.~\ref{fig:graphenedispersion}. In this figure, different graphene dispersions obtained from different models are compared. One can observe the results obtained from our three-parameter non-orthogonal model (blue-dashed line), the two-parameter orthogonal model of Ref.~\onlinecite{CastroNeto2009} (green-dashed line), and an orthogonal model where the first $15$ hopping parameters of Ref.~\onlinecite{Jung2013} are used (red-dashed line).  The figure also depicts the energy dispersion from a first-principle calculation of graphene that was made using the QuantumWise software (black-solid line).\cite{Brandbyge2002}  From the figure, we can observe that the $15$ parameter orthogonal basis model reproduces very well the dispersion relation obtained by first-principle calculations. The two-parameter orthogonal and three-parameter non-orthogonal models capture the essential features, but differ markedly at the $M$ point for the chosen parameters. This is not surprising, as it has been shown that the behaviour around the $M$ point is strongly influenced by higher-order hoppings.\cite{Bena2011}

\subsection{Lattice-preserving bending}
\begin{figure}[tb!]
\centering
\includegraphics[width=0.45\textwidth]{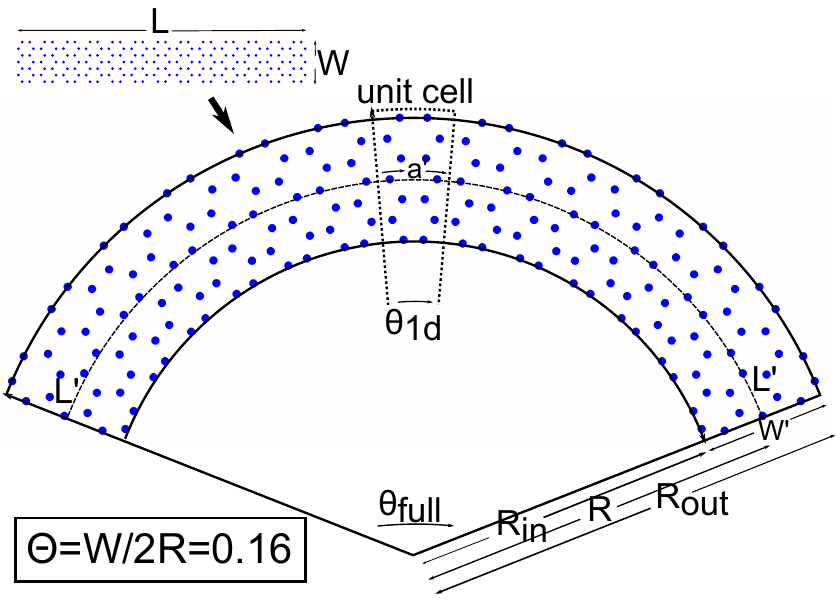}
     \caption{(Color online.) Parameters characterizing the straight and the strained GNR. The parameters $L$ and $W$ characterize the straight ribbon, while the inner and outer radius $R_{in}$ and $R_{out}$ specify perfectly circular bending. Here, we also define the bending radius $R$ along the center of the ribbon, the width of the ribbon after bending $W'=R_{out}-R_{in}$, the length $L'$ along the center of the bent ribbon, the length of a $1$D unit vector $a'$ along the center of the bent unit cell, the curvature $\theta_{1d}$ of a unit cell, the total curvature $\theta_{\rm full}$ of the ribbon and the bending parameter $\Theta=W/2R$. The region enclosed by the dotted line is the $1D$ unit cell of the bent ribbon.}
    \label{fig:bendingconventions}
\end{figure}
To find a minimal model that can describe the geometry of bent GNRs, we first introduce the concept of lattice-preserving bending. This type of deformation can be described by the parameters defined in Fig.~\ref{fig:bendingconventions}. We quantify the degree of bending using the dimensionless parameter $\Theta$, defined as $\Theta=W/2R$, where $W$ is the width of the undistorted ribbon and $R$ is the radius of the circular deformation. For $W'\approx W$, this is approximately equal to the parameter used in Ref.~\onlinecite{Koskinen2012}. For straight ribbons, we can define a $1$D unit cell with sites labelled by $m$, given by $\{\mathbf{r}^{1d}_m\}$, and a $1$D lattice vector $\mathbf{a}$. All sites can then be decomposed in $\mathbf{r}_i=\mathbf{r}^{1d}_m+{\ell} \mathbf{a}$ for some number $\ell$.  This allows us to reduce the size of the matrices in Eq.~(\ref{eq:TBmatrix}) using the $1$D Bloch's theorem to $2N\times2N$, with $N$ the number of dimer lines of the ribbon (number of sublattice pairs in the unit cell, which is always even for ZGNRs), see for instance Ref.~\onlinecite{Son2006}. However, the $1$D translational symmetry that allows this procedure is broken after bending. A lattice-preserving bending is a type of bending deformation that still allows us to reduce the matrices in Eq.~(\ref{eq:TBmatrix}) to size $2N\times2N$. This is possible because a lattice-preserving bending $\mathbf{F}_{\Theta}$ satisfies the discrete rotational symmetry
\begin{equation} \label{eq:discreterotsym}
\mathbf{F}_{\Theta}(\mathbf{r}_i+\mathbf{a})=\mathcal{R}_{-\theta_{1d}}\mathbf{F}_{\Theta}(\mathbf{r}_i),
\end{equation}
where $\mathcal{R}_{-\theta_{1d}}$ is the matrix that represents a clockwise rotation by angle $\theta_{1d}$, and $\mathbf{r}_i$ and $\mathbf{a}$ are the lattice sites and $1$D lattice vector of the straight ribbon, respectively. This symmetry can be seen as a type of modified periodic boundary condition.\cite{Koskinen2012} Because the Hamiltonian commutes with the rotation operator by an angle $\theta_{1d}$, we write a $1D$ Bloch-type wavefunction for a bent GNR in terms of a continuous quantum number $k$. In real space, Eq.~(\ref{eq:TBwave}) then assumes the form
\begin{equation}\label{eq:fqbending}
\psi^{\Theta,k}(\mathbf{r}) = \sum_{\ell ,m} e^{i \ell k} c_{m}^{\Theta,k} \phi(\mathbf{r}-\mathcal{R}_{- \ell \theta_{1d}}\mathbf{r}'^{1d}_m)).
\end{equation}
Here, $k\in[0,2\pi]$, $m$ runs over the atoms in the bent unit cell $\mathbf{r}'^{1d}_m=\mathbf{F}_{\Theta}(\mathbf{r}^{1d}_m)$, and $\ell$ runs over the number of unit cells in the ribbon. The vector $\mathbf{c}^{\Theta,k}=(c^{\Theta,k}_1,\ldots,c^{\Theta,k}_{2N})^T$ therefore completely determines the electron state for a certain wavevector $k$ and bending parameter $\Theta$. Namely, the components $c_j$ of Eq.~(\ref{eq:TBwave}) are given by $c_j=e^{i \ell k} c_{m}^{\Theta,k}$, with $j$ related to $\ell,m$ such that $\mathbf{r}_j=\mathbf{r}^{1d}_m+{\ell} \mathbf{a}$. From the time-independent Schr\"odinger equation~(\ref{eq:TBmatrix}), we can derive a matrix equation for the vector of orbital components $\mathbf{c}^{\Theta,k}$,
\begin{equation}\label{eq:bendingTBmatrix}
(S^{\Theta,k})^{-1} T^{\Theta,k}c^{\Theta,k}=(E_k^{\Theta}-\epsilon_0)\mathbf{c}^{\Theta,k}.
\end{equation}
Here, $T_k$ and $S_k$ are $2N\times2N$ matrices with components
\begin{equation}
\begin{split}
S^{\Theta,k}_{m n} = \sum_{\ell} e^{i k \ell} s\big(\mathcal{R}_{-\ell \theta_{1d}}\mathbf{r}'^{1d}_m - \mathbf{r}'^{1d}_n\big),\\
T^{\Theta,k}_{m n} = \sum_{\ell} e^{i k \ell} t\big(\mathcal{R}_{-\ell \theta_{1d}}\mathbf{r}'^{1d}_m - \mathbf{r}'^{1d}_n\big),
\end{split}
\end{equation}
where $t$ and $s$ are defined as in Eq.~(\ref{eq:exppar}) and $E_k^{\Theta}$ is the spectrum of the eigenstates. In our calculations, we use the values for $s_0$ and $\kappa$ derived from graphene. The on-site energy is set to zero, giving a Fermi level close to, but not exactly at zero. After the calculation, the spectra are shifted by an amount $\epsilon_0$ to place the Fermi level at zero. As can be seen from Eq.~(\ref{eq:bendingTBmatrix}), the dispersion scales linearly with $t_0$ when the scale is normalized around the Fermi level, and we can thus calculate the dispersion in terms of $t_0$ without having to explicitly specify its value. The tight-binding model using non-orthogonal basis and exponentially decaying hopping and overlap in combination with lattice-preserving bending may be  used as a minimal model to study bent GNRs because it only requires three parameters and equations with matrices of size $2N\times2N$.
\subsection{Two types of bending}
A realistic geometry for a bent GNR may be extracted from a molecular dynamics simulation, where bending affects both bond lengths and bond angles. The exact type of bending then depends on the ratio of the spring constants of the respective deformations. From previous work, it is known that the bond length in the graphene lattice is much stiffer than the bond angle.\cite{Li2003,Zhao2011} This observation prompts us to explore a limiting scenario, where bending is completely absorbed in bond-angle distortions, and which we call \emph{bondlength-preserving} bending. In addition, we consider a distortion which we  call \emph{width-preserving} bending, where the atomic positions are rotated around a concentric point. The width-preserving bending is the same deformation as has been used in Ref. \onlinecite{Guinea2010}. Notice that the bond length-preserving bending obeys the rules of graphene kirigami, since the paper can be folded (bond-angle deformations), but it cannot be strained (bond-length deformations). The fixing of the bond lengths in the bond length-preserving bending leaves the NN hopping unchanged, so that any perturbation in the electronic structure can mainly be ascribed to modifications of the NNN hopping. In contrast, bond lengths are allowed to change in the width-preserving bending scheme, so it may be expected that the changes in the dispersion are mainly due to changes in NN hoppings. Comparing the effects of these two types of bending on the spectrum, therefore, allows us to decouple the effects of NN and NNN distortions.

\begin{figure}[tb!]
\centering
\includegraphics[width=0.45\textwidth]{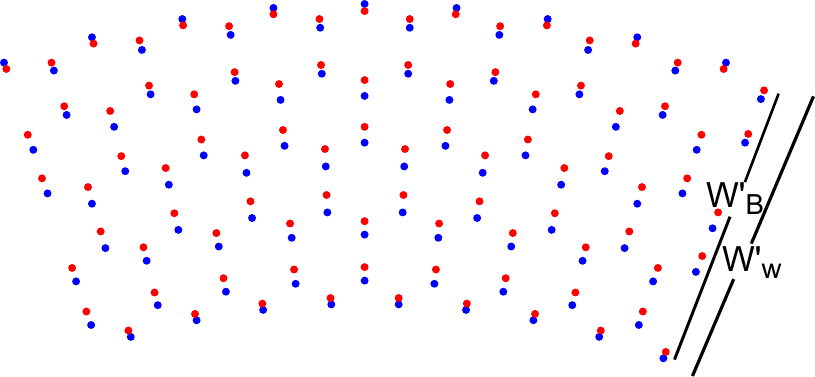}
\caption{(Color online.) A bent ZGNR with $N=5$ and $\Theta=0.15$ for
bond length-preserving (red dots) and width preserving (blue dots). The width of the ribbon after bond length-preserving bending is $W'_B$ and after width-preserving bending is $W'_W$.}
    \label{fig:twobendingways}
\end{figure}

Both bending deformations are depicted in Fig.~\ref{fig:twobendingways}. We can explicitly describe the width-preserving bending by the deformation function
\begin{equation}\label{eq:widthpreserving}
\mathbf{F}^w(\mathbf{r}, \Theta)  = (r_y + R) \left(\begin{array}{c} \sin(r_x/R)  \\ \cos(r_x/R) \end{array}\right),
\end{equation}
with $\mathbf{r}=(r_x,r_y)$. This deformation assumes that the ribbon is positioned such that the middle of the GNR is on the $x$-axis and the ribbon lies in the $xy$-plane. Hence, the $y$ coordinate of the undeformed site is in the interval $[-W/2,W/2]$. One can easily verify that this bending satisfies the definition of a lattice-preserving bending $\mathbf{F}^w(\mathbf{r}+\mathbf{a}_{1d}, \Theta)=\mathcal{R}_{-\theta_{1d}}\mathbf{F}^w(\mathbf{r}, \Theta)$. This deformation is a width-preserving bending in the sense that the distances between sites in the direction along the width of the GNR remain unchanged. Another feature of this bending is that the strain in the direction along the ribbon width increases linearly from the inner to the outer edge. This, in conjunction with the fact that the bending considered here equally compresses on the inside as it stretches outside, yields a line of zero stress exactly in the middle of the ribbon.

It is not straightforward to give a closed formula for the bond length-preserving bending. However, we can construct the profile of the deformation by applying $\mathbf{F}^{NN}(\mathbf{R}_i, \Theta)$ on specific ribbon sites $\mathbf{R}_i$ recursively, see Appendix. The bondlength-preserving bending is similar to the width-preserving one, but has a non-linear strain profile from the bottom to the top of the ribbon. At the inner edge, the ribbon experiences not only longitudinal compressive strain, but also transverse tensile strain. At the outer edge, on the other hand, a compressive transverse strain is present. It is also important to note that the total width becomes reduced, as can be seen in Fig.~\ref{fig:twobendingways}. This reduction of width needs to be taken into account when comparing effects of the bond length-preserving with the width-preserving bending. As a consequence of the reduction of width, the longitudinal strains at the inner ($\epsilon_{in}$) and outer edge ($\epsilon_{out}$) are not identical for the two types of bending.

\section{\label{sec:results}Results: Tunable edge state dispersion}

\begin{figure}[b]
\centering
\includegraphics[width=0.45\textwidth]{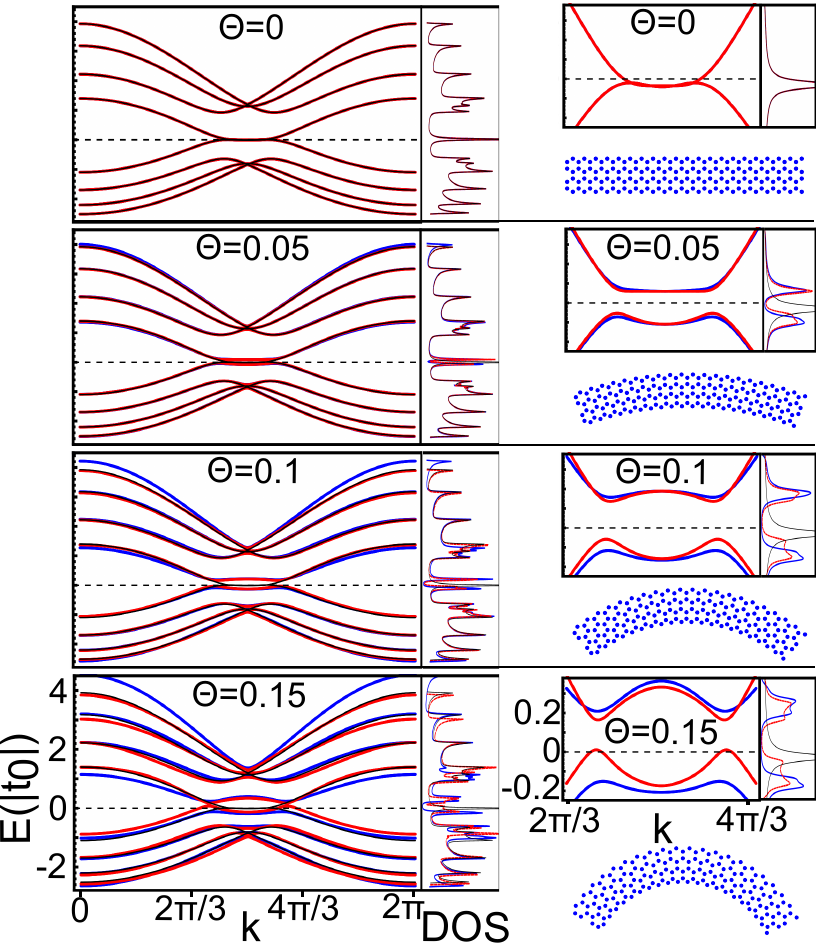}
    \caption{(Color online.) Dispersion relation and DOS for a $N=4$ ZGNR as a function of the bending parameter $\Theta$ for width-preserving bending (blue) and bond length-preserving bending (red). The thin-black line corresponds to a straight ribbon. On the left side, we show the spectrum over the complete $BZ$, for $\Theta$ varying from $0$ at the top to $0.15$ at the bottom panel, with steps of $0.05$. On the right side, we zoom in on the edge state with $k$ ranging from $2\pi/3$ to $4\pi/3$ and depict the lattice of the GNR. All plots have the same scale as shown in the bottom. Calculations were made using non-orthogonal parameters with exponential decay, given by Eq.~(\ref{eq:exppar}),  with $s_0=0.2$ , $\kappa=2.6$, and $\epsilon_0$ such that the Fermi energy (dotted line) of the straight ribbon lies at zero. The DOS is calculated using a Lorentzian broadening with a width of $0.03$ eV (DOS in arbitrary units).}\label{fig:EkbentZGNR}
\end{figure}

\begin{figure*}[tbh!]
\centering
\includegraphics[width=\textwidth]{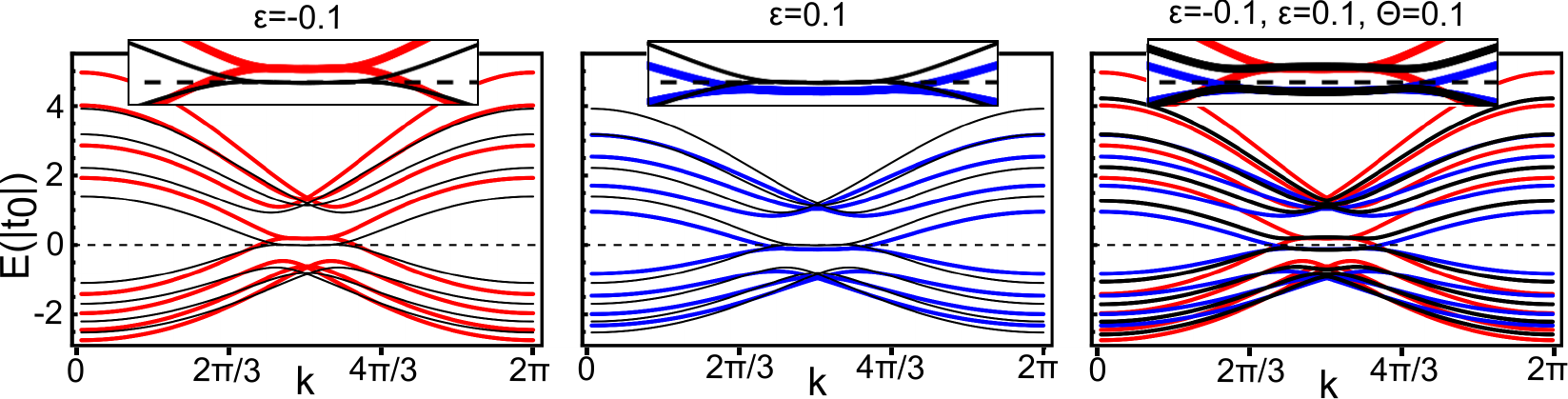}
    \caption{(Color online.) Dispersion relation for a $N=4$ ZGNR for uniform strain $\epsilon=-0.1$ (left panel, red line) and $\epsilon=0.1$ (middle panel, blue line). The thinner-black line in the left and middle picture corresponds to the straight ribbon. The right panel shows the dispersions for uniform strain  $\epsilon=-0.1$ (red), $\epsilon=0.1$ (blue), and width-preserving bending with $\Theta=0.1$ (black). Calculations were performed using non-orthogonal parameters with exponential decay, given by Eq.~(\ref{eq:exppar}) with $s_0=0.2$ , $\kappa=2.6$, and $\epsilon_0$ such that the Fermi energy (dotted line) of the straight ribbon lies at zero.}\label{fig:EkstretchZGNR}
\end{figure*}

We have calculated the dispersion relation for bent ZGNRs by solving Eq.~(\ref{eq:bendingTBmatrix}) numerically both for width-preserving and for bond length-preserving bending. In Fig.~\ref{fig:EkbentZGNR},  the dispersion relation for different values of the bending parameter is depicted. Since we argued that bending introduces a profile of elastic deformation with effective compressive strain on the inside and tensile strain on the outside, it is useful to compare it to the effects of a uniform longitudinal strain $\epsilon$, defined as $\epsilon=\Delta L/L$, where $\Delta L$ is the length deformation introduced by the strain, and $L$ is the length of the undeformed nanoribbon. Fig.~\ref{fig:EkstretchZGNR} depicts the effect of positive (tensile) and negative (compressive) longitudinal strain on a $N=4$ ZGNR subjected to a width-preserving uniform strain deformation. We can see that the energy of the edge states increases (decreases) for negative (positive) strain. When we compare the two cases with a ribbon bent using width-preserving bending, we observe that the energy increase in the edge state that experiences compression is roughly equal to the energy increase in both edge states of a longitudinally compressed ribbon. Similarly, we find a good agreement for the outer edge state with both edge states of a ribbon experiencing tensile strain. These observations indicate that the dispersion of ribbons bent by $\Theta$ is quantitatively related to the dispersion of a uniformly strained ribbon with strain $\epsilon=\pm\Theta$, a result consistent with Ref.~\onlinecite{Koskinen2012}.

\begin{figure}[bh!]
\centering
\includegraphics[width=0.45\textwidth]{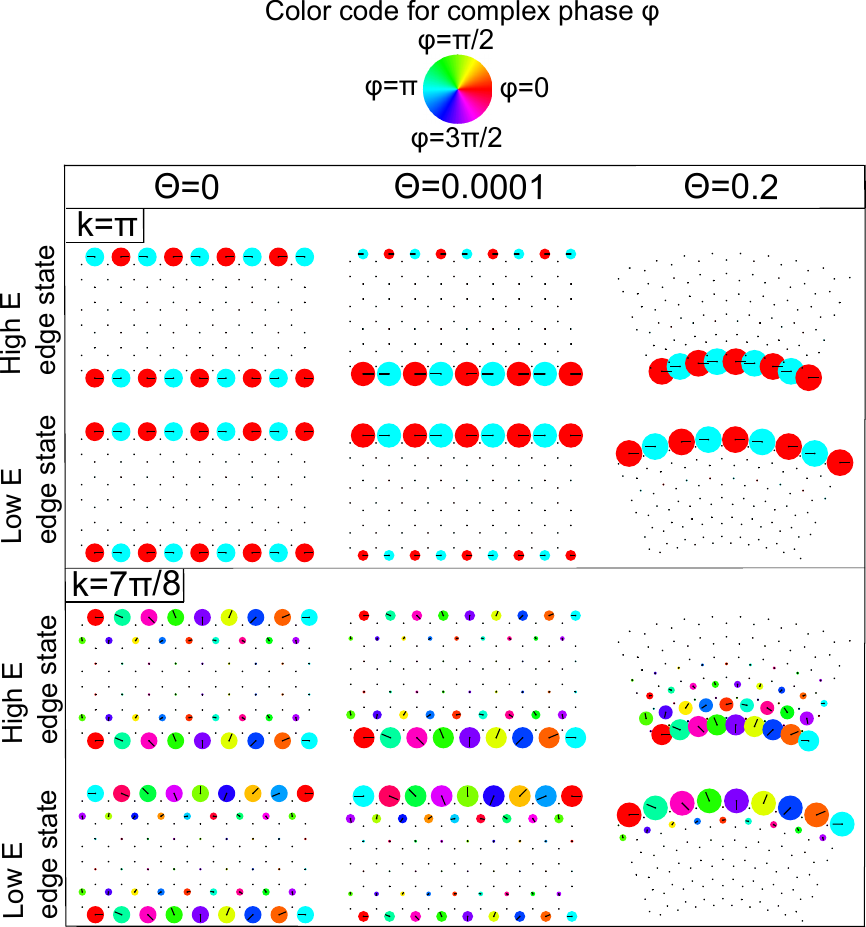}
    \caption{(Color online.) Edge states in the real space. Components $c_{\ell,m}$ of edge states for a section of the ribbon at $k=\pi$ and $k=7\pi/8$ are mapped to the corresponding $\mathbf{r}_i$ points of the bent GNR for different values of bending-parameter $\Theta$. The width-preserving bending scheme was used here in combination with our standard orbital hopping parameters. $c_{\ell,m}$ are related to the eigenvector through the definition Eq.~($\ref{eq:fqbending}$) and satisfy Eq.~(\ref{eq:bendingTBmatrix}). The coefficients $c_{\ell,m}$ are complex numbers that are depicted in two ways. The first is by dots of which the diameter is proportional to the absolute value of $c_{\ell,m}$ and the color corresponds to the phase, as indicated by the color code. Additionally  the coefficients $c_{\ell,m}$ are represented by a vector in the complex plane (see small  black lines at the center of the dots). The phase is chosen such that the lattice site at the left bottom of the picture has phase zero. The top (bottom) rows of the $k=\pi$ and $k=7\pi/8$ panels correspond to the high-energy (low-energy) edge states. The ribbon has width $N=6$.}\label{fig:edgestatesbent}
\end{figure}

Plotting the wavefunctions of the edge states confirms that the low-energy state resides on the outside, as shown in Fig.~\ref{fig:edgestatesbent}. Here, the orbital components of the eigenstates of the edge states, $c_j$, as defined in Eq.~($\ref{eq:fqbending}$), are plotted for increasing $\Theta$. 
The width-preserving bending scheme was used in generating the plots. First, we note that the edge states are localized on one sublattice at both edges, forming a symmetric and antisymmetric combination of states localized on either edge. The states are almost degenerate, which would allow us to form orthonormal combinations that are still eigenstates of the Hamiltonian with the same eigenenergy. In fact, since interaction effects arising from the Coulomb repulsion  are not  accounted for, we may expect these effects to favor a different combination in the two-dimensional Hilbert space of eigenstates. Intuitively, the effect of electron-electron repulsion should be to split the symmetric and antisymmetric states into two states that are localized on their respective edges, both singly occupied.

With increasing bending, we observe that the nearly degenerate states that initially reside on both edges in our model transform into a high-energy state localized on the inner edge and a low-energy state localized on the outer edge. It is interesting to note that this already occurs for the very small bending parameter of $\Theta=10^{-4}$, indicating that for this strength of bending, the  symmetric and antisymmetric states mix in order to form the   states localized on a single edge, energetically more favorable. We also find a significant dependence of the localization length of both edge states on the momentum $k$. When we plot, for example, the edge states for a wave vector of $k=7\pi/8$, the wave function appears to spread more into the bulk of the ribbon than for the value $k=\pi$, as shown in Fig.~\ref{fig:edgestatesbent}. Although not shown here, the results for bond length-preserving bending show that the edge state for $k=7\pi/8$ is also less localized than for $k=\pi$. However, for the same degree of bending, the effect is much less pronounced than for width-preserving bending.

\begin{figure}[bh!]
\centering
\includegraphics[width=0.4\textwidth]{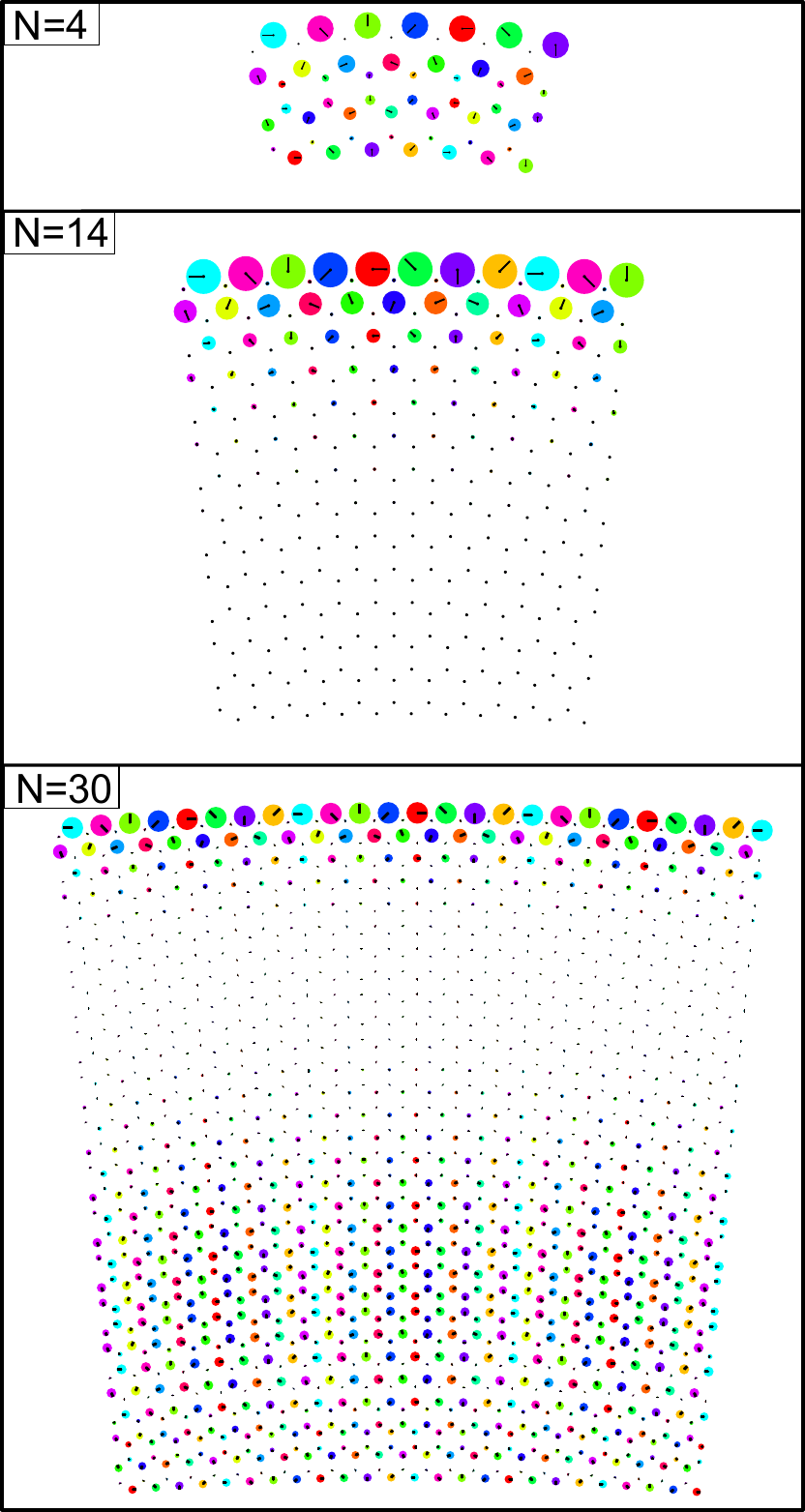}
    \caption{(Color online.) Real space depiction of lower energy edge states for ribbons of with $N=4$, $N=14$, $N=30$. Construction is the same as explained in Fig.~\ref{fig:edgestatesbent}. For all three ribbons the bending parameter is $\Theta=0.1$ and $k=6\pi/8$.  The $N=4$ ribbon shows no clear edge states because the states are hybridized across the entire ribbon, while the $N=30$ ribbon shows localized edge states, which are, however, hybridized with ones in the bulk .}\label{fig:ribbonwidth}
\end{figure}
Another striking observation is that the two edge states do not only split but also develop opposite curvature, as shown in Fig.~\ref{fig:EkbentZGNR}. The top band is curved upward, but at its center a small downward curvature develops, such that there is a local maximum at $k=\pi$, whereas the opposite occurs for the lower band. This is in contrast with what we observe for positive or negative uniform strain in Fig.~\ref{fig:EkstretchZGNR}. In that case, the edge states are only shifted, but retain the same dispersion as in the strain-free ribbon.

A minimal model that captures this behaviour, and in particular fits the dispersion of the edge states around the point $k=\pi$, is a tight-binding model of a $1$D chain of sites with a single NN hopping $t^{h/l}$ and an on-site energy $\epsilon^{h/l}$. Here, the superscripts refer to the higher-energy band and lower-energy band, which are localized on the inner and outer edge, respectively. The effective dispersion obtained from the $1$D NN tight-binding model reads
\begin{equation}\label{eq:dispersion1d}
E^{h/l}_{k}(\Theta)=\epsilon_0^{h/l}(\Theta)+2t^{h/l}(\Theta)\cos(k).
\end{equation}
Inspection of the zoomed in panels of Fig.~\ref{fig:EkbentZGNR} suggest that this effective model can describe the shape of the bands in the region around $k=\pi$ reasonably well.  A positive or negative $t^{h/l}$ relates to the dispersion that exhibits, respectively, an upwards or downwards curvature around momentum $k=\pi$.

Before we compare this effective model quantitatively with the tight-binding results, we need to mention the effect of the width of the ribbon on the edge states. As a ribbon becomes narrower, the edge state localized on one side with $k$ {\it closer} to $k=\pi$ starts to hybridize with the edge state localized on the other edge.  On the other hand, when one starts bending a ribbon the edge states start moving closer in energy to the bulk states. This can be seen in Fig.~\ref{fig:EkbentZGNR}. After a certain bending, the valence band maximum hybridizes with the lower-energy edge states, as well as the conduction band minimum hybridizes with the higher-energy edge states. Since wider ribbons have a smaller bulk band gap, these effects are more pronounced. These effects are shown in Fig.~\ref{fig:ribbonwidth}, where we plot the lower-energy edge state for three different widths of the ribbon, $N=4,\, 14,$ and 30, using the same bending parameter, $\Theta=0.1$, and $k-$value  $k=6\pi/8$.  We observe that the two edge states  of the $N = 4$ ribbon  hybridize with each other, and are therefore not localized anymore. The edge states of the ribbon with $N=30$ also hybridize, but instead with bulk states, and are  not localized anymore either. The ribbon with $N=14$, however, still shows  localized edge states for the same regime of parameters. These two opposite effects make the comparison between different ribbon sizes very intricate. We have chosen to analyze the $N=14$ ribbon in more detail because this one has the optimal width to avoid spurious hybridization effects of the first or second kind. Our observations are expected to hold also for ribbons of different width, if care is taken to account for these hybridization effects.

\begin{figure*}[tbh!]
\includegraphics[width=\textwidth]{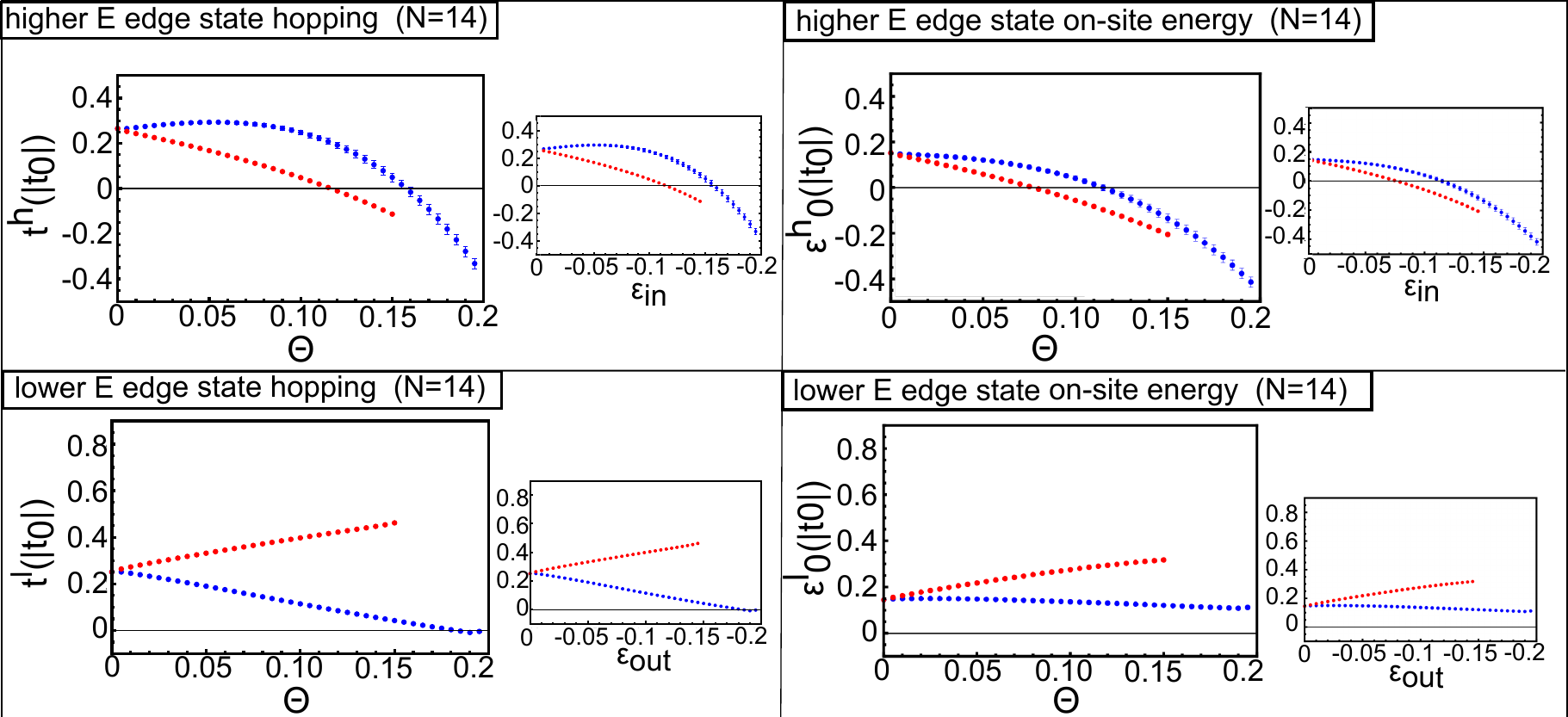}
\caption{(Color online.) Best fit values of the effective $1$D chain parameters $t^{h}$, $\epsilon_0^{h}$ (upper rectangular panel) and $t^{l}$, $\epsilon_0^{l}$ (lower rectangular panel) versus the bending parameter $\Theta$. These parameters are defined in Eq.~(\ref{eq:dispersion1d}). The fitting has been performed for width-preserving bending (blue dots) and bond length-preserving bending (red dots). The right picture in each panel shows these parameters with respect to the strain on the inner edge $\epsilon_{in}$ for the higher-energy edge state and with respect to the strain on the outer edge $\epsilon_{out}$ for the lower-energy edge states. The relation between $\epsilon_{in}$, $\epsilon_{out}$ and $\Theta$ is explained in the text. The fitting was  done for a ZGNR of width $N=14$, and is based on data points chosen in the region around $k=\pi$ given by $k\in[2.41,3.86]$. Error bars are obtained from the standard deviation between the fitted spectra and the numerics on the lattice. All calculations are performed using our standard set of exponentially decaying orbital hopping parameters.}\label{fig:1dfit}
\end{figure*}

We fit the parameters of the effective $1$D dispersion of Eq.~\ref{eq:dispersion1d} to the tight-binding calculations for a ribbon of width $N=14$. In Fig.~\ref{fig:1dfit}, we plot the fitted parameters for different values of the bending $\Theta$. We observe that both the lower- and the higher-energy edge states start out with the same positive hopping parameter. Interestingly, in both bending schemes, $t^h$ crosses zero, implying that for a certain bending parameter the band becomes dispersionless. This is an important observation because many-body effects can be expected to become even more relevant for that bending parameter.

By comparing how the parameters change with respect to the type of bending used  we can identify whether the NN or the NNN hopping is more important. The effective parameters for the state on the inner edge decrease for both types of bending. However, for the outer edge the effective parameters increase for bondlength-preserving bending, but decrease for width-preserving bending. The main difference between the two bending methods is that in the width-preserving bending also the NN distance is modified.  Therefore, we can conclude that for the outer-edge state NN effects are more important than for the inner-edge. General behavior of the inner-edge state, however, can be captured by only considering the effect of the NNN hopping. If we compare the effective parameters for the inner-edge state between the two bending methods in more detail, we observe that the effective parameters for bond length-preserving bending show a linear dependence on $\Theta$, while this dependence for width-preserving bending is more complicated. One reason for this behavior could be the fact that the width- and bond length-preserving bending produce a small difference in strain on the edges ($\epsilon_{in}$, $\epsilon_{out}$). To check whether this can account for the difference, we also plot the effective parameters as a function of the strain (smaller plots in Fig.~\ref{fig:1dfit}). We can clearly see that the general behavior does not change. Therefore, the difference should be sought in effects of the NN hopping. Changes in the NN distance influences the hybridization between the opposite edges and the hybridization of the edge state with bulk states. These effects might explain why the effective parameters of width-preserving bending exhibit a nonlinear dependence on the bending. Furthermore, the effect of the perturbation of the NN distance also depends on the width of the ribbon, which additionally complicates the problem. Because of all this, in the following we focus only on the effective parameters of bond-length-preserving bending.

For bond length-preserving bending (see plots in red in Fig.~\ref{fig:1dfit}), the effective hopping at the inner edge (higher-$E$) linearly decreases and changes sign, whereas the hopping at the outer edge (lower-$E$) linearly increases. We could try to understand this behaviour by assuming a perfectly localized edge state. The inner edge experiences a negative strain, so the hopping becomes more negative and the $1$D dispersion would curve downwards. This indeed corresponds to what we observe in Fig.~\ref{fig:1dfit}. On the same token,  the hopping at the outer edge should decrease, because the distances between the lattice sites increase, and therefore a flat band should develop. However, the opposite behaviour is visible in Fig.~\ref{fig:1dfit}. This can be understood by noting that the changes due to bending at the outer edge are determined by the weight of the wavefunction on sites closer to the bulk. This is because these sites are closer to each other and therefore contribute more to the energy. This together with the fact that sites close to the bulk have a sizeable weight implies that our assumption of the localized states does not apply. The fact that the edge state becomes less localized as the momentum moves further away from $k=\pi$ is crucial here. This enhances the effect that can already be seen for straight ribbons, where the edge states are dispersive at the momenta away from $k=\pi$,  and causes an increasing positive effective hopping.

In conclusion, we can understand the behaviour as a competition between two effects due to NNN hopping and strain:
 \begin{enumerate}
   \item An effective positive hopping for increasing negative strain because of the increasing delocalized nature of the edge state as the momentum moves further away from $k=\pi$.
   \item An effective negative hopping for increasing negative strain because the edge state is localized.
 \end{enumerate}
For the edge state localized on the outer edge, the first effect is always dominant and becomes even more relevant after bending. For the inner edge, the second effect overcomes the first after a certain bending parameter. This is the reason why the dispersion of the inner edge has to go through a point at which it is dispersionless. This also clarifies our earlier observation that the outer edge state is more sensitive to changes in the NN hopping. 
The outer edges are more delocalized, and therefore more sensitive to the effects of the NN hoppings.

\section{\label{sec:conclusions}Conclusions}
We show here that a tight-binding model with exponentially decaying hopping and overlap can be used as a minimal model with three parameters to study a graphene-based system that is both geometrically confined and strained. To obtain geometries of bent nanoribbons that serve as the input of the tight-binding model, we develop two types of bending, bond length-preserving and width-preserving. We would like to point out that bond-length preserving bending geometry, generated using a recursive algorithm, shows a particularly strong analogy with the Japanese art of \emph{kirigami}. Both types of bending are lattice-preserving, causing the resulting geometry to exhibit rotational symmetry (the unit cell is rotated by $\theta_{1d}$ to generate the entire bent GNR), and therefore allowing us to reduce the tight-binding model to the numerically inexpensive problem of solving a matrix equation with $2N\times2N$ matrices, with $2N$ the number of sites in the unit cell of the GNR. The different types of bending allow us to decouple the effects of perturbations of the NN and NNN parameters of the tight-binding model.

We have investigated the qualitative features of the dispersion relation upon bending. Our calculations show that bending leads to nontrivial effects on the edge states of ZGNRs, resulting from the broken symmetry between the top and bottom edges. We observe that both width-preserving and bond length-preserving bending predict a splitting of the two edge states (without considering interactions). A lower-energy edge state localizes on the outer edge and a higher-energy edge state on the inner edge. In fact, there is an emergent band structure around the point $k=\pi$ of the edge states that can be fitted to the tight-binding dispersion of a $1$D chain with an effective hopping and on-site energy parameter. The higher-energy edge state has an effective hopping parameter that changes sign as the bending is cranked up from $\Theta=0.11$ to $\Theta=0.17$, with the exact value where the effective hopping vanishes depending on the type of bending. Hence, there is a critical degree of bending at which the band is effectively flat and interaction effects are expected to become prominent. Since the charge carrier velocity is proportional to the slope of the dispersion, the degree of localization of the edge states can be tuned with bending. By comparing the two bending methods, we can conclude that effects on the dispersion of the inner-edge state are dominated by changes in NNN hopping. For the outer edge state, changes in NN hopping also become important. The effects due to NN hopping changes, however, are less universal and depend on width and bending method. The effects of the NNN hopping on the emergent band structure at the edges can be explained by a competition between the decreasing localization of the electronic states with the momenta away from $k=\pi$ and the localized character of the edge state. A next step would be to include interaction effects, as these are important for edge states, especially when the bending gives rise to the flat bands. Furthermore, motivated by our work, it would be important to understand how bending would affect the magnetic polarization of the edge states detected recently.\cite{Magda2014} We hope that our results will stimulate further research in these directions.

\section{Acknowledgments.}

We are grateful to Ingmar Swart for fruitful discussions on electronic structure theory, graphene nanoribbons, and their interplay with scanning probe microscopy experiments.

\section*{Appendix}

{\bf Recursion formula for bond length-preserving bending of ZGNR}\label{app:bond length}\\
We construct the bond length-preserving bending, $\mathbf{F}^{NN}$, for a ZGNR. First, we construct the bent $1$D unit cell. The orientation is chosen such that the first site in the bent unit cell is positioned at $\mathbf{r}'^{1d}_1=( 0  , R-W'/2 )$. Note that we do not know $W'$ and $R$ yet, but they will be obtained using a recursive procedure outlined below. We can now recursively generate the next atoms in the deformed $1$D unit cell using the following rule:
\begin{equation*}
\mathbf{r}'^{1d}_m = \begin{cases}  f(\mathbf{r}'^{1d}_{m-1},\theta_{1d}) \mathcal{R}_{-\theta_{1d}/2} \widehat{\mathbf{r}}'^{1d}_{m-1}  & \mathrm{if\ } i \mathrm{\ is\ even}\\
               (|\mathbf{r}'^{1d}_{m-1}| + a) \widehat{\mathbf{r}}'^{1d}_{m-1}       & \mathrm{if\ } i \mathrm{\ is\ odd},\\
           \end{cases}
\end{equation*}
\begin{equation*}
\begin{split}
&f(\mathbf{r}'^{1d}_{m-1},\theta_{1d}) = |\mathbf{r}'^{1d}_{m-1}|\cos(\theta_{1d}/2)\\
&+\sqrt{(a)^2-|\mathbf{r}'^{1d}_{m-1}|^2\sin^2(\theta_{1d}/2)}.
\end{split}
\end{equation*}
Here, $\widehat{\mathbf{r}}'^{1d}_{m-1}$ is the unit vector in the direction of $\mathbf{r}'^{1d}_{m-1}$. We still assume that the distance along the middle of the GNR remains unchanged, and therefore $\theta_{1d}=\Theta a'/(W/2)$. If we follow this recursion until $\mathbf{r}'^{1d}_{2 N}$, where $N$ is the number of $A$ sites in the $1$D unit cell, we have generated the deformed $1$D unit cell $\mathbf{r}'^{1d}_m$. However, we started with $\mathbf{r}'^{1d}_1$ defined in terms of the bent GNR width $W'$, which was unknown. We can now use the identity $W'=|\mathbf{r}'^{1d}_{2N}|-|\mathbf{r}'^{1d}_1|$, which is an equation with $W'$ on both sides, to write out the recursion explicitly. However, this is a rather involved equation. We can, on the other hand, easily find a good approximation iteratively for $W'$. We start with the assumption that $W'\approx W$. Then, after running the recursion, we calculate the $W'$ of that ribbon. If it differs by more than a set test value from the previous recursion, we use that value of $W'$ to generate a new unit cell. This iterative procedure runs until the test condition, that gives the minimal difference between a new and old width, is satisfied.  Note also that this deformation does not work for every $\Theta$, as for large enough bending the square root in the definition will become complex. This is understandable, as there should be a maximum bending at which the lattice sites on the outer edge of the ribbon are all separated by $a$.
Once the bent unit cell is generated, the complete bent GNR is obtained by copying the unit cell through multiples of rotations by $\theta_{1d}$. Thus, we can describe the bond length-preserving bending as
\begin{equation*}\label{eq:bondpreserving}
\mathbf{F}^{NN}(\mathbf{r}^{1d}_m + \ell \mathbf{a}, \Theta) = \mathcal{R}_{-\ell\theta_{1d}}\mathbf{r}'^{1d}_m.
\end{equation*}
Here, $\ell$ runs over the number of unit cells in the ribbon. We explicitly use that the lattice sites of a straight GNR can be described by a site in the $1$D unit cell plus a multiple of $\mathbf{a}$, the lattice vector of the straight ribbon. One can show, using simple trigonometry, that each site now has $3$ neighbors that are at a distance equal to $a$, as shown in Fig.~\ref{fig:twobendingways}. Due to the construction, it is obvious that the rotational symmetry is satisfied and thus this is a lattice-preserving bending.

\addcontentsline{toc}{section}{Bibliography}
\bibliographystyle{apsrev4-1}
\bibliography{Revised-PRB-final}

\begin{thebibliography}{36}%
\makeatletter
\providecommand \@ifxundefined [1]{%
 \@ifx{#1\undefined}
}%
\providecommand \@ifnum [1]{%
 \ifnum #1\expandafter \@firstoftwo
 \else \expandafter \@secondoftwo
 \fi
}%
\providecommand \@ifx [1]{%
 \ifx #1\expandafter \@firstoftwo
 \else \expandafter \@secondoftwo
 \fi
}%
\providecommand \natexlab [1]{#1}%
\providecommand \enquote  [1]{``#1''}%
\providecommand \bibnamefont  [1]{#1}%
\providecommand \bibfnamefont [1]{#1}%
\providecommand \citenamefont [1]{#1}%
\providecommand \href@noop [0]{\@secondoftwo}%
\providecommand \href [0]{\begingroup \@sanitize@url \@href}%
\providecommand \@href[1]{\@@startlink{#1}\@@href}%
\providecommand \@@href[1]{\endgroup#1\@@endlink}%
\providecommand \@sanitize@url [0]{\catcode `\\12\catcode `\$12\catcode
  `\&12\catcode `\#12\catcode `\^12\catcode `\_12\catcode `\%12\relax}%
\providecommand \@@startlink[1]{}%
\providecommand \@@endlink[0]{}%
\providecommand \url  [0]{\begingroup\@sanitize@url \@url }%
\providecommand \@url [1]{\endgroup\@href {#1}{\urlprefix }}%
\providecommand \urlprefix  [0]{URL }%
\providecommand \Eprint [0]{\href }%
\providecommand \doibase [0]{http://dx.doi.org/}%
\providecommand \selectlanguage [0]{\@gobble}%
\providecommand \bibinfo  [0]{\@secondoftwo}%
\providecommand \bibfield  [0]{\@secondoftwo}%
\providecommand \translation [1]{[#1]}%
\providecommand \BibitemOpen [0]{}%
\providecommand \bibitemStop [0]{}%
\providecommand \bibitemNoStop [0]{.\EOS\space}%
\providecommand \EOS [0]{\spacefactor3000\relax}%
\providecommand \BibitemShut  [1]{\csname bibitem#1\endcsname}%
\let\auto@bib@innerbib\@empty
\bibitem [{\citenamefont {Castro~Neto}\ \emph {et~al.}(2009)\citenamefont
  {Castro~Neto}, \citenamefont {Guinea}, \citenamefont {Peres}, \citenamefont
  {Novoselov},\ and\ \citenamefont {Geim}}]{CastroNeto2009}%
  \BibitemOpen
  \bibfield  {author} {\bibinfo {author} {\bibfnamefont {A.~H.}\ \bibnamefont
  {Castro~Neto}}, \bibinfo {author} {\bibfnamefont {F.}~\bibnamefont {Guinea}},
  \bibinfo {author} {\bibfnamefont {N.~M.~R.}\ \bibnamefont {Peres}}, \bibinfo
  {author} {\bibfnamefont {K.~S.}\ \bibnamefont {Novoselov}}, \ and\ \bibinfo
  {author} {\bibfnamefont {A.~K.}\ \bibnamefont {Geim}},\ }\href {\doibase
  10.1103/RevModPhys.81.109} {\bibfield  {journal} {\bibinfo  {journal} {Rev.
  Mod. Phys.}\ }\textbf {\bibinfo {volume} {81}},\ \bibinfo {pages} {109}
  (\bibinfo {year} {2009})}\BibitemShut {NoStop}%
\bibitem [{\citenamefont {Geim}\ and\ \citenamefont
  {Novoselov}(2007)}]{Geim2007}%
  \BibitemOpen
  \bibfield  {author} {\bibinfo {author} {\bibfnamefont {A.~K.}\ \bibnamefont
  {Geim}}\ and\ \bibinfo {author} {\bibfnamefont {K.~S.}\ \bibnamefont
  {Novoselov}},\ }\href@noop {} {\bibfield  {journal} {\bibinfo  {journal}
  {Nat. Mater.}\ }\textbf {\bibinfo {volume} {6}},\ \bibinfo {pages} {183}
  (\bibinfo {year} {2007})}\BibitemShut {NoStop}%
\bibitem [{\citenamefont {Wang}\ \emph {et~al.}(2008)\citenamefont {Wang},
  \citenamefont {Ouyang}, \citenamefont {Li}, \citenamefont {Wang},
  \citenamefont {Guo},\ and\ \citenamefont {Dai}}]{Wang2008}%
  \BibitemOpen
  \bibfield  {author} {\bibinfo {author} {\bibfnamefont {X.}~\bibnamefont
  {Wang}}, \bibinfo {author} {\bibfnamefont {Y.}~\bibnamefont {Ouyang}},
  \bibinfo {author} {\bibfnamefont {X.}~\bibnamefont {Li}}, \bibinfo {author}
  {\bibfnamefont {H.}~\bibnamefont {Wang}}, \bibinfo {author} {\bibfnamefont
  {J.}~\bibnamefont {Guo}}, \ and\ \bibinfo {author} {\bibfnamefont
  {H.}~\bibnamefont {Dai}},\ }\href {\doibase 10.1103/PhysRevLett.100.206803}
  {\bibfield  {journal} {\bibinfo  {journal} {Phys. Rev. Lett.}\ }\textbf
  {\bibinfo {volume} {100}},\ \bibinfo {pages} {206803} (\bibinfo {year}
  {2008})}\BibitemShut {NoStop}%
\bibitem [{\citenamefont {Bennett}\ \emph {et~al.}(2013)\citenamefont
  {Bennett}, \citenamefont {Pedramrazi}, \citenamefont {Madani}, \citenamefont
  {Chen}, \citenamefont {de~Oteyza}, \citenamefont {Chen}, \citenamefont
  {Fischer}, \citenamefont {Crommie},\ and\ \citenamefont
  {Bokor}}]{Bennett2013}%
  \BibitemOpen
  \bibfield  {author} {\bibinfo {author} {\bibfnamefont {P.~B.}\ \bibnamefont
  {Bennett}}, \bibinfo {author} {\bibfnamefont {Z.}~\bibnamefont {Pedramrazi}},
  \bibinfo {author} {\bibfnamefont {A.}~\bibnamefont {Madani}}, \bibinfo
  {author} {\bibfnamefont {Y.-C.}\ \bibnamefont {Chen}}, \bibinfo {author}
  {\bibfnamefont {D.~G.}\ \bibnamefont {de~Oteyza}}, \bibinfo {author}
  {\bibfnamefont {C.}~\bibnamefont {Chen}}, \bibinfo {author} {\bibfnamefont
  {F.~R.}\ \bibnamefont {Fischer}}, \bibinfo {author} {\bibfnamefont {M.~F.}\
  \bibnamefont {Crommie}}, \ and\ \bibinfo {author} {\bibfnamefont
  {J.}~\bibnamefont {Bokor}},\ }\href {\doibase
  http://dx.doi.org/10.1063/1.4855116} {\bibfield  {journal} {\bibinfo
  {journal} {Appl. Phys. Lett.}\ }\textbf {\bibinfo {volume} {103}},\ \bibinfo
  {eid} {253114} (\bibinfo {year} {2013})}\BibitemShut {NoStop}%
\bibitem [{\citenamefont {Fujita}\ \emph {et~al.}(1996)\citenamefont {Fujita},
  \citenamefont {Wakabayashi}, \citenamefont {Nakada},\ and\ \citenamefont
  {Kusakabe}}]{Fujita1996}%
  \BibitemOpen
  \bibfield  {author} {\bibinfo {author} {\bibfnamefont {M.}~\bibnamefont
  {Fujita}}, \bibinfo {author} {\bibfnamefont {K.}~\bibnamefont {Wakabayashi}},
  \bibinfo {author} {\bibfnamefont {K.}~\bibnamefont {Nakada}}, \ and\ \bibinfo
  {author} {\bibfnamefont {K.}~\bibnamefont {Kusakabe}},\ }\href {\doibase
  10.1143/JPSJ.65.1920} {\bibfield  {journal} {\bibinfo  {journal} {J. Phys.
  Soc. Jpn.}\ }\textbf {\bibinfo {volume} {65}},\ \bibinfo {pages} {1920}
  (\bibinfo {year} {1996})}\BibitemShut {NoStop}%
\bibitem [{\citenamefont {Cai}\ \emph {et~al.}(2010)\citenamefont {Cai},
  \citenamefont {Ruffieux}, \citenamefont {Jaafar}, \citenamefont {Bieri},
  \citenamefont {Braun}, \citenamefont {Blankenburg}, \citenamefont {Muoth},
  \citenamefont {Seitsonen}, \citenamefont {Saleh}, \citenamefont {Feng},
  \citenamefont {M\"{u}llen},\ and\ \citenamefont {Fasel}}]{Cai2010}%
  \BibitemOpen
  \bibfield  {author} {\bibinfo {author} {\bibfnamefont {J.}~\bibnamefont
  {Cai}}, \bibinfo {author} {\bibfnamefont {P.}~\bibnamefont {Ruffieux}},
  \bibinfo {author} {\bibfnamefont {R.}~\bibnamefont {Jaafar}}, \bibinfo
  {author} {\bibfnamefont {M.}~\bibnamefont {Bieri}}, \bibinfo {author}
  {\bibfnamefont {T.}~\bibnamefont {Braun}}, \bibinfo {author} {\bibfnamefont
  {S.}~\bibnamefont {Blankenburg}}, \bibinfo {author} {\bibfnamefont
  {M.}~\bibnamefont {Muoth}}, \bibinfo {author} {\bibfnamefont {A.~P.}\
  \bibnamefont {Seitsonen}}, \bibinfo {author} {\bibfnamefont {M.}~\bibnamefont
  {Saleh}}, \bibinfo {author} {\bibfnamefont {X.}~\bibnamefont {Feng}},
  \bibinfo {author} {\bibfnamefont {K.}~\bibnamefont {M\"{u}llen}}, \ and\
  \bibinfo {author} {\bibfnamefont {R.}~\bibnamefont {Fasel}},\ }\href
  {\doibase 10.1038/nature09211} {\bibfield  {journal} {\bibinfo  {journal}
  {Nature (Londen)}\ }\textbf {\bibinfo {volume} {466}},\ \bibinfo {pages}
  {470} (\bibinfo {year} {2010})}\BibitemShut {NoStop}%
\bibitem [{\citenamefont {Chen}\ \emph {et~al.}(2013)\citenamefont {Chen},
  \citenamefont {de~Oteyza}, \citenamefont {Pedramrazi}, \citenamefont {Chen},
  \citenamefont {Fischer},\ and\ \citenamefont {Crommie}}]{Chen2013}%
  \BibitemOpen
  \bibfield  {author} {\bibinfo {author} {\bibfnamefont {Y.-C.}\ \bibnamefont
  {Chen}}, \bibinfo {author} {\bibfnamefont {D.~G.}\ \bibnamefont {de~Oteyza}},
  \bibinfo {author} {\bibfnamefont {Z.}~\bibnamefont {Pedramrazi}}, \bibinfo
  {author} {\bibfnamefont {C.}~\bibnamefont {Chen}}, \bibinfo {author}
  {\bibfnamefont {F.~R.}\ \bibnamefont {Fischer}}, \ and\ \bibinfo {author}
  {\bibfnamefont {M.~F.}\ \bibnamefont {Crommie}},\ }\href {\doibase
  10.1021/nn401948e} {\bibfield  {journal} {\bibinfo  {journal} {ACS Nano}\
  }\textbf {\bibinfo {volume} {7}},\ \bibinfo {pages} {6123} (\bibinfo {year}
  {2013})}\BibitemShut {NoStop}%
\bibitem [{\citenamefont {Narita}\ \emph {et~al.}(2014)\citenamefont {Narita},
  \citenamefont {Feng}, \citenamefont {Hernandez}, \citenamefont {Jensen},
  \citenamefont {Bonn}, \citenamefont {Yang}, \citenamefont {Verzhbitskiy},
  \citenamefont {Casiraghi}, \citenamefont {Hansen}, \citenamefont {Koch},
  \citenamefont {Fytas}, \citenamefont {Ivasenko}, \citenamefont {Li},
  \citenamefont {Mali}, \citenamefont {Balandina}, \citenamefont {Mahesh},
  \citenamefont {De~Feyter},\ and\ \citenamefont {M\"ullen}}]{Narita2014}%
  \BibitemOpen
  \bibfield  {author} {\bibinfo {author} {\bibfnamefont {A.}~\bibnamefont
  {Narita}}, \bibinfo {author} {\bibfnamefont {X.}~\bibnamefont {Feng}},
  \bibinfo {author} {\bibfnamefont {Y.}~\bibnamefont {Hernandez}}, \bibinfo
  {author} {\bibfnamefont {S.~A.}\ \bibnamefont {Jensen}}, \bibinfo {author}
  {\bibfnamefont {M.}~\bibnamefont {Bonn}}, \bibinfo {author} {\bibfnamefont
  {H.}~\bibnamefont {Yang}}, \bibinfo {author} {\bibfnamefont {I.~A.}\
  \bibnamefont {Verzhbitskiy}}, \bibinfo {author} {\bibfnamefont
  {C.}~\bibnamefont {Casiraghi}}, \bibinfo {author} {\bibfnamefont {M.~R.}\
  \bibnamefont {Hansen}}, \bibinfo {author} {\bibfnamefont {A.~H.~R.}\
  \bibnamefont {Koch}}, \bibinfo {author} {\bibfnamefont {G.}~\bibnamefont
  {Fytas}}, \bibinfo {author} {\bibfnamefont {O.}~\bibnamefont {Ivasenko}},
  \bibinfo {author} {\bibfnamefont {B.}~\bibnamefont {Li}}, \bibinfo {author}
  {\bibfnamefont {K.~S.}\ \bibnamefont {Mali}}, \bibinfo {author}
  {\bibfnamefont {T.}~\bibnamefont {Balandina}}, \bibinfo {author}
  {\bibfnamefont {S.}~\bibnamefont {Mahesh}}, \bibinfo {author} {\bibfnamefont
  {S.}~\bibnamefont {De~Feyter}}, \ and\ \bibinfo {author} {\bibfnamefont
  {K.}~\bibnamefont {M\"ullen}},\ }\href {\doibase 10.1038/nchem.1819}
  {\bibfield  {journal} {\bibinfo  {journal} {Nat. Chem.}\ }\textbf {\bibinfo
  {volume} {6}},\ \bibinfo {pages} {126} (\bibinfo {year} {2014})}\BibitemShut
  {NoStop}%
\bibitem [{\citenamefont {Basagni}\ \emph {et~al.}(2015)\citenamefont
  {Basagni}, \citenamefont {Sedona}, \citenamefont {Pignedoli}, \citenamefont
  {Cattelan}, \citenamefont {Nicolas}, \citenamefont {Casarin},\ and\
  \citenamefont {Sambi}}]{Basagni2015}%
  \BibitemOpen
  \bibfield  {author} {\bibinfo {author} {\bibfnamefont {A.}~\bibnamefont
  {Basagni}}, \bibinfo {author} {\bibfnamefont {F.}~\bibnamefont {Sedona}},
  \bibinfo {author} {\bibfnamefont {C.~A.}\ \bibnamefont {Pignedoli}}, \bibinfo
  {author} {\bibfnamefont {M.}~\bibnamefont {Cattelan}}, \bibinfo {author}
  {\bibfnamefont {L.}~\bibnamefont {Nicolas}}, \bibinfo {author} {\bibfnamefont
  {M.}~\bibnamefont {Casarin}}, \ and\ \bibinfo {author} {\bibfnamefont
  {M.}~\bibnamefont {Sambi}},\ }\href {\doibase 10.1021/ja510292b} {\bibfield
  {journal} {\bibinfo  {journal} {J. Am. Chem. Soc.}\ }\textbf {\bibinfo
  {volume} {137}},\ \bibinfo {pages} {1802} (\bibinfo {year}
  {2015})}\BibitemShut {NoStop}%
\bibitem [{\citenamefont {Shi}\ \emph {et~al.}(2011)\citenamefont {Shi},
  \citenamefont {Yang}, \citenamefont {Zhang}, \citenamefont {Wang},
  \citenamefont {Liu}, \citenamefont {Shi}, \citenamefont {Wang},\ and\
  \citenamefont {Zhang}}]{Shi2011}%
  \BibitemOpen
  \bibfield  {author} {\bibinfo {author} {\bibfnamefont {Z.}~\bibnamefont
  {Shi}}, \bibinfo {author} {\bibfnamefont {R.}~\bibnamefont {Yang}}, \bibinfo
  {author} {\bibfnamefont {L.}~\bibnamefont {Zhang}}, \bibinfo {author}
  {\bibfnamefont {Y.}~\bibnamefont {Wang}}, \bibinfo {author} {\bibfnamefont
  {D.}~\bibnamefont {Liu}}, \bibinfo {author} {\bibfnamefont {D.}~\bibnamefont
  {Shi}}, \bibinfo {author} {\bibfnamefont {E.}~\bibnamefont {Wang}}, \ and\
  \bibinfo {author} {\bibfnamefont {G.}~\bibnamefont {Zhang}},\ }\href@noop {}
  {\bibfield  {journal} {\bibinfo  {journal} {Advanced Materials}\ }\textbf
  {\bibinfo {volume} {23}},\ \bibinfo {pages} {3061} (\bibinfo {year}
  {2011})}\BibitemShut {NoStop}%
\bibitem [{\citenamefont {Han}\ \emph {et~al.}(2014)\citenamefont {Han},
  \citenamefont {Akagi}, \citenamefont {Federici~Canova}, \citenamefont
  {Mutoh}, \citenamefont {Shiraki}, \citenamefont {Iwaya}, \citenamefont
  {Weiss}, \citenamefont {Asao},\ and\ \citenamefont {Hitosugi}}]{Han2014}%
  \BibitemOpen
  \bibfield  {author} {\bibinfo {author} {\bibfnamefont {P.}~\bibnamefont
  {Han}}, \bibinfo {author} {\bibfnamefont {K.}~\bibnamefont {Akagi}}, \bibinfo
  {author} {\bibfnamefont {F.}~\bibnamefont {Federici~Canova}}, \bibinfo
  {author} {\bibfnamefont {H.}~\bibnamefont {Mutoh}}, \bibinfo {author}
  {\bibfnamefont {S.}~\bibnamefont {Shiraki}}, \bibinfo {author} {\bibfnamefont
  {K.}~\bibnamefont {Iwaya}}, \bibinfo {author} {\bibfnamefont {P.~S.}\
  \bibnamefont {Weiss}}, \bibinfo {author} {\bibfnamefont {N.}~\bibnamefont
  {Asao}}, \ and\ \bibinfo {author} {\bibfnamefont {T.}~\bibnamefont
  {Hitosugi}},\ }\href {\doibase 10.1021/nn5028642} {\bibfield  {journal}
  {\bibinfo  {journal} {ACS Nano}\ }\textbf {\bibinfo {volume} {8}},\ \bibinfo
  {pages} {9181} (\bibinfo {year} {2014})}\BibitemShut {NoStop}%
\bibitem [{\citenamefont {Magda}\ \emph {et~al.}(2014)\citenamefont {Magda},
  \citenamefont {Jin}, \citenamefont {Hagym{\'a}si}, \citenamefont
  {Vancs{\'o}}, \citenamefont {Osv{\'a}th}, \citenamefont {Nemes-Incze},
  \citenamefont {Hwang}, \citenamefont {Bir{\'o}},\ and\ \citenamefont
  {Tapaszt{\'o}}}]{Magda2014}%
  \BibitemOpen
  \bibfield  {author} {\bibinfo {author} {\bibfnamefont {G.}~\bibnamefont
  {Magda}}, \bibinfo {author} {\bibfnamefont {X.}~\bibnamefont {Jin}}, \bibinfo
  {author} {\bibfnamefont {I.}~\bibnamefont {Hagym{\'a}si}}, \bibinfo {author}
  {\bibfnamefont {P.}~\bibnamefont {Vancs{\'o}}}, \bibinfo {author}
  {\bibfnamefont {Z.}~\bibnamefont {Osv{\'a}th}}, \bibinfo {author}
  {\bibfnamefont {P.}~\bibnamefont {Nemes-Incze}}, \bibinfo {author}
  {\bibfnamefont {C.}~\bibnamefont {Hwang}}, \bibinfo {author} {\bibfnamefont
  {L.~P.}\ \bibnamefont {Bir{\'o}}}, \ and\ \bibinfo {author} {\bibfnamefont
  {L.}~\bibnamefont {Tapaszt{\'o}}},\ }\href@noop {} {\bibfield  {journal}
  {\bibinfo  {journal} {Nature (Londen)}\ }\textbf {\bibinfo {volume} {514}},\
  \bibinfo {pages} {608} (\bibinfo {year} {2014})}\BibitemShut {NoStop}%
\bibitem [{\citenamefont {Son}\ \emph {et~al.}(2006)\citenamefont {Son},
  \citenamefont {Cohen},\ and\ \citenamefont {Louie}}]{Son2006}%
  \BibitemOpen
  \bibfield  {author} {\bibinfo {author} {\bibfnamefont {Y.}~\bibnamefont
  {Son}}, \bibinfo {author} {\bibfnamefont {M.}~\bibnamefont {Cohen}}, \ and\
  \bibinfo {author} {\bibfnamefont {S.}~\bibnamefont {Louie}},\ }\href
  {\doibase 10.1103/PhysRevLett.97.216803} {\bibfield  {journal} {\bibinfo
  {journal} {Phys. Rev. Lett.}\ }\textbf {\bibinfo {volume} {97}},\ \bibinfo
  {pages} {216803} (\bibinfo {year} {2006})}\BibitemShut {NoStop}%
\bibitem [{\citenamefont {Kane}\ and\ \citenamefont {Mele}(1997)}]{Kane1997}%
  \BibitemOpen
  \bibfield  {author} {\bibinfo {author} {\bibfnamefont {C.~L.}\ \bibnamefont
  {Kane}}\ and\ \bibinfo {author} {\bibfnamefont {E.~J.}\ \bibnamefont
  {Mele}},\ }\href {\doibase 10.1103/PhysRevLett.78.1932} {\bibfield  {journal}
  {\bibinfo  {journal} {Phys. Rev. Lett.}\ }\textbf {\bibinfo {volume} {78}},\
  \bibinfo {pages} {1932} (\bibinfo {year} {1997})}\BibitemShut {NoStop}%
\bibitem [{\citenamefont {Amorim}\ \emph {et~al.}()\citenamefont {Amorim},
  \citenamefont {Cortijo}, \citenamefont {de~Juan}, \citenamefont {Grushin},
  \citenamefont {Guinea}, \citenamefont {Gutiérrez-Rubio}, \citenamefont
  {Ochoa}, \citenamefont {Parente}, \citenamefont {Roldán}, \citenamefont
  {San-José}, \citenamefont {Schiefele}, \citenamefont {Sturla},\ and\
  \citenamefont {Vozmediano}}]{Amorim2015}%
  \BibitemOpen
  \bibfield  {author} {\bibinfo {author} {\bibfnamefont {B.}~\bibnamefont
  {Amorim}}, \bibinfo {author} {\bibfnamefont {A.}~\bibnamefont {Cortijo}},
  \bibinfo {author} {\bibfnamefont {F.}~\bibnamefont {de~Juan}}, \bibinfo
  {author} {\bibfnamefont {A.~G.}\ \bibnamefont {Grushin}}, \bibinfo {author}
  {\bibfnamefont {F.}~\bibnamefont {Guinea}}, \bibinfo {author} {\bibfnamefont
  {A.}~\bibnamefont {Gutiérrez-Rubio}}, \bibinfo {author} {\bibfnamefont
  {H.}~\bibnamefont {Ochoa}}, \bibinfo {author} {\bibfnamefont
  {V.}~\bibnamefont {Parente}}, \bibinfo {author} {\bibfnamefont
  {R.}~\bibnamefont {Roldán}}, \bibinfo {author} {\bibfnamefont
  {P.}~\bibnamefont {San-José}}, \bibinfo {author} {\bibfnamefont
  {J.}~\bibnamefont {Schiefele}}, \bibinfo {author} {\bibfnamefont
  {M.}~\bibnamefont {Sturla}}, \ and\ \bibinfo {author} {\bibfnamefont
  {M.~A.~H.}\ \bibnamefont {Vozmediano}},\ }\href@noop {} {\ }\Eprint
  {http://arxiv.org/abs/1503.00747} {arXiv:1503.00747} \BibitemShut {NoStop}%
\bibitem [{\citenamefont {Pereira}\ and\ \citenamefont
  {Castro~Neto}(2009)}]{Pereira2009a}%
  \BibitemOpen
  \bibfield  {author} {\bibinfo {author} {\bibfnamefont {V.~M.}\ \bibnamefont
  {Pereira}}\ and\ \bibinfo {author} {\bibfnamefont {A.~H.}\ \bibnamefont
  {Castro~Neto}},\ }\href {\doibase 10.1103/PhysRevLett.103.046801} {\bibfield
  {journal} {\bibinfo  {journal} {Phys. Rev. Lett.}\ }\textbf {\bibinfo
  {volume} {103}},\ \bibinfo {pages} {046801} (\bibinfo {year}
  {2009})}\BibitemShut {NoStop}%
\bibitem [{\citenamefont {Levy}\ \emph {et~al.}(2010)\citenamefont {Levy},
  \citenamefont {Burke}, \citenamefont {Meaker}, \citenamefont {Panlasigui},
  \citenamefont {Zettl}, \citenamefont {Guinea}, \citenamefont {Neto},\ and\
  \citenamefont {Crommie}}]{Levy2010}%
  \BibitemOpen
  \bibfield  {author} {\bibinfo {author} {\bibfnamefont {N.}~\bibnamefont
  {Levy}}, \bibinfo {author} {\bibfnamefont {S.~A.}\ \bibnamefont {Burke}},
  \bibinfo {author} {\bibfnamefont {K.~L.}\ \bibnamefont {Meaker}}, \bibinfo
  {author} {\bibfnamefont {M.}~\bibnamefont {Panlasigui}}, \bibinfo {author}
  {\bibfnamefont {A.}~\bibnamefont {Zettl}}, \bibinfo {author} {\bibfnamefont
  {F.}~\bibnamefont {Guinea}}, \bibinfo {author} {\bibfnamefont {A.~H.~C.}\
  \bibnamefont {Neto}}, \ and\ \bibinfo {author} {\bibfnamefont {M.~F.}\
  \bibnamefont {Crommie}},\ }\href {\doibase 10.1126/science.1191700}
  {\bibfield  {journal} {\bibinfo  {journal} {Science}\ }\textbf {\bibinfo
  {volume} {329}},\ \bibinfo {pages} {544} (\bibinfo {year}
  {2010})}\BibitemShut {NoStop}%
\bibitem [{\citenamefont {Avdoshenko}\ \emph {et~al.}(2013)\citenamefont
  {Avdoshenko}, \citenamefont {Koskinen}, \citenamefont {Sevin{\c{c}}li},
  \citenamefont {Popov},\ and\ \citenamefont {Rocha}}]{Avdoshenko2013}%
  \BibitemOpen
  \bibfield  {author} {\bibinfo {author} {\bibfnamefont {S.~M.}\ \bibnamefont
  {Avdoshenko}}, \bibinfo {author} {\bibfnamefont {P.}~\bibnamefont
  {Koskinen}}, \bibinfo {author} {\bibfnamefont {H.}~\bibnamefont
  {Sevin{\c{c}}li}}, \bibinfo {author} {\bibfnamefont {A.~A.}\ \bibnamefont
  {Popov}}, \ and\ \bibinfo {author} {\bibfnamefont {C.~G.}\ \bibnamefont
  {Rocha}},\ }\href@noop {} {\bibfield  {journal} {\bibinfo  {journal} {Sci.
  Rep.}\ }\textbf {\bibinfo {volume} {3}} (\bibinfo {year} {2013})}\BibitemShut
  {NoStop}%
\bibitem [{\citenamefont {Korhonen}\ and\ \citenamefont
  {Koskinen}(2014)}]{Korhonen2014}%
  \BibitemOpen
  \bibfield  {author} {\bibinfo {author} {\bibfnamefont {T.}~\bibnamefont
  {Korhonen}}\ and\ \bibinfo {author} {\bibfnamefont {P.}~\bibnamefont
  {Koskinen}},\ }\href@noop {} {\bibfield  {journal} {\bibinfo  {journal} {AIP
  Adv.}\ }\textbf {\bibinfo {volume} {4}},\ \bibinfo {pages} {127125} (\bibinfo
  {year} {2014})}\BibitemShut {NoStop}%
\bibitem [{\citenamefont {Zhang}\ and\ \citenamefont {Zhao}(2014)}]{Zhang2014}%
  \BibitemOpen
  \bibfield  {author} {\bibinfo {author} {\bibfnamefont {X.}~\bibnamefont
  {Zhang}}\ and\ \bibinfo {author} {\bibfnamefont {M.}~\bibnamefont {Zhao}},\
  }\href@noop {} {\bibfield  {journal} {\bibinfo  {journal} {Sci. Rep.}\
  }\textbf {\bibinfo {volume} {4}} (\bibinfo {year} {2014})}\BibitemShut
  {NoStop}%
\bibitem [{\citenamefont {Koskinen}(2012)}]{Koskinen2012}%
  \BibitemOpen
  \bibfield  {author} {\bibinfo {author} {\bibfnamefont {P.}~\bibnamefont
  {Koskinen}},\ }\href {\doibase 10.1103/PhysRevB.85.205429} {\bibfield
  {journal} {\bibinfo  {journal} {Phys. Rev. B}\ }\textbf {\bibinfo {volume}
  {85}},\ \bibinfo {pages} {205429} (\bibinfo {year} {2012})}\BibitemShut
  {NoStop}%
\bibitem [{\citenamefont {Guinea}\ \emph {et~al.}(2010)\citenamefont {Guinea},
  \citenamefont {Geim}, \citenamefont {Katsnelson},\ and\ \citenamefont
  {Novoselov}}]{Guinea2010}%
  \BibitemOpen
  \bibfield  {author} {\bibinfo {author} {\bibfnamefont {F.}~\bibnamefont
  {Guinea}}, \bibinfo {author} {\bibfnamefont {A.~K.}\ \bibnamefont {Geim}},
  \bibinfo {author} {\bibfnamefont {M.~I.}\ \bibnamefont {Katsnelson}}, \ and\
  \bibinfo {author} {\bibfnamefont {K.~S.}\ \bibnamefont {Novoselov}},\ }\href
  {\doibase 10.1103/PhysRevB.81.035408} {\bibfield  {journal} {\bibinfo
  {journal} {Phys. Rev. B}\ }\textbf {\bibinfo {volume} {81}},\ \bibinfo
  {pages} {035408} (\bibinfo {year} {2010})}\BibitemShut {NoStop}%
\bibitem [{\citenamefont {van~der Lit}\ \emph {et~al.}(2015)\citenamefont
  {van~der Lit}, \citenamefont {Jacobse}, \citenamefont {Vanmaekelbergh},\ and\
  \citenamefont {Swart}}]{Lit2015}%
  \BibitemOpen
  \bibfield  {author} {\bibinfo {author} {\bibfnamefont {J.}~\bibnamefont
  {van~der Lit}}, \bibinfo {author} {\bibfnamefont {P.~H.}\ \bibnamefont
  {Jacobse}}, \bibinfo {author} {\bibfnamefont {D.~A.~M.}\ \bibnamefont
  {Vanmaekelbergh}}, \ and\ \bibinfo {author} {\bibfnamefont {I.}~\bibnamefont
  {Swart}},\ }\href@noop {} {\bibfield  {journal} {\bibinfo  {journal} {New J.
  Phys.}\ }\textbf {\bibinfo {volume} {17}},\ \bibinfo {pages} {053013}
  (\bibinfo {year} {2015})}\BibitemShut {NoStop}%
\bibitem [{\citenamefont {Qi}\ \emph {et~al.}(2014)\citenamefont {Qi},
  \citenamefont {Campbell},\ and\ \citenamefont {Park}}]{Qi2014}%
  \BibitemOpen
  \bibfield  {author} {\bibinfo {author} {\bibfnamefont {Z.}~\bibnamefont
  {Qi}}, \bibinfo {author} {\bibfnamefont {D.~K.}\ \bibnamefont {Campbell}}, \
  and\ \bibinfo {author} {\bibfnamefont {H.~S.}\ \bibnamefont {Park}},\ }\href
  {\doibase 10.1103/PhysRevB.90.245437} {\bibfield  {journal} {\bibinfo
  {journal} {Phys. Rev. B}\ }\textbf {\bibinfo {volume} {90}},\ \bibinfo
  {pages} {245437} (\bibinfo {year} {2014})}\BibitemShut {NoStop}%
\bibitem [{\citenamefont {Goerbig}(2011)}]{Goerbig2011}%
  \BibitemOpen
  \bibfield  {author} {\bibinfo {author} {\bibfnamefont {M.~O.}\ \bibnamefont
  {Goerbig}},\ }\href {\doibase 10.1103/RevModPhys.83.1193} {\bibfield
  {journal} {\bibinfo  {journal} {Rev. Mod. Phys.}\ }\textbf {\bibinfo {volume}
  {83}},\ \bibinfo {pages} {1193} (\bibinfo {year} {2011})}\BibitemShut
  {NoStop}%
\bibitem [{\citenamefont {Pereira}\ \emph {et~al.}(2009)\citenamefont
  {Pereira}, \citenamefont {Castro~Neto},\ and\ \citenamefont
  {Peres}}]{Pereira2009b}%
  \BibitemOpen
  \bibfield  {author} {\bibinfo {author} {\bibfnamefont {V.~M.}\ \bibnamefont
  {Pereira}}, \bibinfo {author} {\bibfnamefont {A.~H.}\ \bibnamefont
  {Castro~Neto}}, \ and\ \bibinfo {author} {\bibfnamefont {N.~M.~R.}\
  \bibnamefont {Peres}},\ }\href {\doibase 10.1103/PhysRevB.80.045401}
  {\bibfield  {journal} {\bibinfo  {journal} {Phys. Rev. B}\ }\textbf {\bibinfo
  {volume} {80}},\ \bibinfo {pages} {045401} (\bibinfo {year}
  {2009})}\BibitemShut {NoStop}%
\bibitem [{\citenamefont {Ribeiro}\ \emph {et~al.}(2009)\citenamefont
  {Ribeiro}, \citenamefont {Pereira}, \citenamefont {Peres}, \citenamefont
  {Briddon},\ and\ \citenamefont {Castro~Neto}}]{Ribeiro2009}%
  \BibitemOpen
  \bibfield  {author} {\bibinfo {author} {\bibfnamefont {R.~M.}\ \bibnamefont
  {Ribeiro}}, \bibinfo {author} {\bibfnamefont {V.~M.}\ \bibnamefont
  {Pereira}}, \bibinfo {author} {\bibfnamefont {N.~M.~R.}\ \bibnamefont
  {Peres}}, \bibinfo {author} {\bibfnamefont {P.~R.}\ \bibnamefont {Briddon}},
  \ and\ \bibinfo {author} {\bibfnamefont {A.~H.}\ \bibnamefont
  {Castro~Neto}},\ }\href {\doibase 10.1088/1367-2630/11/11/115002} {\bibfield
  {journal} {\bibinfo  {journal} {New J. Phys.}\ }\textbf {\bibinfo {volume}
  {11}},\ \bibinfo {pages} {115002} (\bibinfo {year} {2009})}\BibitemShut
  {NoStop}%
\bibitem [{\citenamefont {Jung}\ and\ \citenamefont
  {MacDonald}(2013)}]{Jung2013}%
  \BibitemOpen
  \bibfield  {author} {\bibinfo {author} {\bibfnamefont {J.}~\bibnamefont
  {Jung}}\ and\ \bibinfo {author} {\bibfnamefont {A.~H.}\ \bibnamefont
  {MacDonald}},\ }\href {\doibase 10.1103/PhysRevB.87.195450} {\bibfield
  {journal} {\bibinfo  {journal} {Phys. Rev. B}\ }\textbf {\bibinfo {volume}
  {87}},\ \bibinfo {pages} {195450} (\bibinfo {year} {2013})}\BibitemShut
  {NoStop}%
\bibitem [{\citenamefont {Reich}\ \emph {et~al.}(2002)\citenamefont {Reich},
  \citenamefont {Maultzsch}, \citenamefont {Thomsen},\ and\ \citenamefont
  {Ordej\'on}}]{Reich2002}%
  \BibitemOpen
  \bibfield  {author} {\bibinfo {author} {\bibfnamefont {S.}~\bibnamefont
  {Reich}}, \bibinfo {author} {\bibfnamefont {J.}~\bibnamefont {Maultzsch}},
  \bibinfo {author} {\bibfnamefont {C.}~\bibnamefont {Thomsen}}, \ and\
  \bibinfo {author} {\bibfnamefont {P.}~\bibnamefont {Ordej\'on}},\ }\href
  {\doibase 10.1103/PhysRevB.66.035412} {\bibfield  {journal} {\bibinfo
  {journal} {Phys. Rev. B}\ }\textbf {\bibinfo {volume} {66}},\ \bibinfo
  {pages} {035412} (\bibinfo {year} {2002})}\BibitemShut {NoStop}%
\bibitem [{\citenamefont {Aiken}\ \emph {et~al.}(1980)\citenamefont {Aiken},
  \citenamefont {Erdos},\ and\ \citenamefont {Goldstein}}]{Aiken1980}%
  \BibitemOpen
  \bibfield  {author} {\bibinfo {author} {\bibfnamefont {J.~G.}\ \bibnamefont
  {Aiken}}, \bibinfo {author} {\bibfnamefont {J.~A.}\ \bibnamefont {Erdos}}, \
  and\ \bibinfo {author} {\bibfnamefont {J.~A.}\ \bibnamefont {Goldstein}},\
  }\href {\doibase 10.1002/qua.560180416} {\bibfield  {journal} {\bibinfo
  {journal} {Int. J. Quant. Chem.}\ }\textbf {\bibinfo {volume} {18}},\
  \bibinfo {pages} {1101} (\bibinfo {year} {1980})}\BibitemShut {NoStop}%
\bibitem [{\citenamefont {Sasaki}\ \emph {et~al.}(2009)\citenamefont {Sasaki},
  \citenamefont {Shimomura}, \citenamefont {Takane},\ and\ \citenamefont
  {Wakabayashi}}]{Sasaki2009}%
  \BibitemOpen
  \bibfield  {author} {\bibinfo {author} {\bibfnamefont {K.}~\bibnamefont
  {Sasaki}}, \bibinfo {author} {\bibfnamefont {Y.}~\bibnamefont {Shimomura}},
  \bibinfo {author} {\bibfnamefont {Y.}~\bibnamefont {Takane}}, \ and\ \bibinfo
  {author} {\bibfnamefont {K.}~\bibnamefont {Wakabayashi}},\ }\href {\doibase
  10.1103/PhysRevLett.102.146806} {\bibfield  {journal} {\bibinfo  {journal}
  {Phys. Rev. Lett.}\ }\textbf {\bibinfo {volume} {102}},\ \bibinfo {pages}
  {146806} (\bibinfo {year} {2009})}\BibitemShut {NoStop}%
\bibitem [{\citenamefont {Hancock}\ \emph {et~al.}(2010)\citenamefont
  {Hancock}, \citenamefont {Uppstu}, \citenamefont {Saloriutta}, \citenamefont
  {Harju},\ and\ \citenamefont {Puska}}]{Hancock2010}%
  \BibitemOpen
  \bibfield  {author} {\bibinfo {author} {\bibfnamefont {Y.}~\bibnamefont
  {Hancock}}, \bibinfo {author} {\bibfnamefont {A.}~\bibnamefont {Uppstu}},
  \bibinfo {author} {\bibfnamefont {K.}~\bibnamefont {Saloriutta}}, \bibinfo
  {author} {\bibfnamefont {A.}~\bibnamefont {Harju}}, \ and\ \bibinfo {author}
  {\bibfnamefont {M.~J.}\ \bibnamefont {Puska}},\ }\href {\doibase
  10.1103/PhysRevB.81.245402} {\bibfield  {journal} {\bibinfo  {journal} {Phys.
  Rev. B}\ }\textbf {\bibinfo {volume} {81}},\ \bibinfo {pages} {245402}
  (\bibinfo {year} {2010})}\BibitemShut {NoStop}%
\bibitem [{\citenamefont {Brandbyge}\ \emph {et~al.}(2002)\citenamefont
  {Brandbyge}, \citenamefont {Mozos}, \citenamefont {Ordej\'on}, \citenamefont
  {Taylor},\ and\ \citenamefont {Stokbro}}]{Brandbyge2002}%
  \BibitemOpen
  \bibfield  {author} {\bibinfo {author} {\bibfnamefont {M.}~\bibnamefont
  {Brandbyge}}, \bibinfo {author} {\bibfnamefont {J.}~\bibnamefont {Mozos}},
  \bibinfo {author} {\bibfnamefont {P.}~\bibnamefont {Ordej\'on}}, \bibinfo
  {author} {\bibfnamefont {J.}~\bibnamefont {Taylor}}, \ and\ \bibinfo {author}
  {\bibfnamefont {K.}~\bibnamefont {Stokbro}},\ }\href {\doibase
  10.1103/PhysRevB.65.165401} {\bibfield  {journal} {\bibinfo  {journal} {Phys.
  Rev. B}\ }\textbf {\bibinfo {volume} {65}},\ \bibinfo {pages} {165401}
  (\bibinfo {year} {2002})}\BibitemShut {NoStop}%
\bibitem [{\citenamefont {Bena}\ and\ \citenamefont {Simon}(2011)}]{Bena2011}%
  \BibitemOpen
  \bibfield  {author} {\bibinfo {author} {\bibfnamefont {C.}~\bibnamefont
  {Bena}}\ and\ \bibinfo {author} {\bibfnamefont {L.}~\bibnamefont {Simon}},\
  }\href {\doibase 10.1103/PhysRevB.83.115404} {\bibfield  {journal} {\bibinfo
  {journal} {Phys. Rev. B}\ }\textbf {\bibinfo {volume} {83}},\ \bibinfo
  {pages} {115404} (\bibinfo {year} {2011})}\BibitemShut {NoStop}%
\bibitem [{\citenamefont {Li}\ and\ \citenamefont {Chou}(2003)}]{Li2003}%
  \BibitemOpen
  \bibfield  {author} {\bibinfo {author} {\bibfnamefont {C.}~\bibnamefont
  {Li}}\ and\ \bibinfo {author} {\bibfnamefont {T.-W.}\ \bibnamefont {Chou}},\
  }\href {\doibase 10.1016/S0020-7683(03)00056-8} {\bibfield  {journal}
  {\bibinfo  {journal} {Int. J. Solids Struct.}\ }\textbf {\bibinfo {volume}
  {40}},\ \bibinfo {pages} {2487} (\bibinfo {year} {2003})}\BibitemShut
  {NoStop}%
\bibitem [{\citenamefont {Zhao}\ and\ \citenamefont {Shi}(2011)}]{Zhao2011}%
  \BibitemOpen
  \bibfield  {author} {\bibinfo {author} {\bibfnamefont {P.}~\bibnamefont
  {Zhao}}\ and\ \bibinfo {author} {\bibfnamefont {G.}~\bibnamefont {Shi}},\
  }\href@noop {} {\bibfield  {journal} {\bibinfo  {journal} {Tech Science SL.}\
  }\textbf {\bibinfo {volume} {5}},\ \bibinfo {pages} {49} (\bibinfo {year}
  {2011})}\BibitemShut {NoStop}%
\end{thebibliography}%

\end{document}